\documentclass[a4paper,12pt]{article}

\usepackage{epsfig}
\usepackage{a4}
\usepackage{latexsym, amssymb} 

\newcommand{\lapprox}{%
\mathrel{%
\setbox0=\hbox{$<$}
\raise0.6ex\copy0\kern-\wd0
\lower0.65ex\hbox{$\sim$}
}}
\newcommand{\gapprox}{%
\mathrel{%
\setbox0=\hbox{$>$}
\raise0.6ex\copy0\kern-\wd0
\lower0.65ex\hbox{$\sim$}
}}

\newcommand{\ba}{\begin{array}}
\newcommand{\ea}{\end{array}}
\newcommand{\bd}{\begin{displaymath}}
\newcommand{\ed}{\end{displaymath}}
\newcommand{\be}{\begin{equation}}
\newcommand{\ee}{\end{equation}}
\newcommand{\bea}{\begin{eqnarray}}
\newcommand{\eea}{\end{eqnarray}}

\def\fb{\, {\rm fb}}
\def\pb{\, {\rm pb}}
\def\met{E_T \hspace*{-1.1em}/\hspace*{0.5em}}
\def\tev{\, \, {\rm TeV}}
\def\gev{\, \, {\rm GeV}}

\newcommand{\BR}{\mathrm{BR}}

\newcommand{\PYTHIA}{PYTHIA}

\def\order#1{{\cal O}\left(#1\right)}

\def\thefootnote{\fnsymbol{footnote}}

\begin{document}

\begin{titlepage}

\begin{flushright}
{\small 
March 28, 2010 \\ 
RECAPP-HRI-2009-022}
\end{flushright}

\vspace*{0.2cm}
\begin{center}
{\Large {\bf Dilepton and Four-Lepton Signals at the LHC \\[0.2cm]
in the Littlest Higgs Model with T-parity Violation}}\\[2cm]  
Satyanarayan Mukhopadhyay\footnote{satya@hri.res.in}, 
Biswarup Mukhopadhyaya\footnote{biswarup@hri.res.in}, 
Andreas Nyf\/feler\footnote{nyf\/feler@hri.res.in} \\[0.5cm]  

{\it Regional Centre for Accelerator-based Particle Physics \\
Harish-Chandra Research Institute \\
Chhatnag Road, Jhusi \\ 
Allahabad - 211 019, India}\\[2cm]
\end{center}

\begin{abstract}
In the presence of the T-parity violating Wess-Zumino-Witten (WZW) anomaly
term, the otherwise stable heavy photon $A_H$ in the Littlest Higgs model with
T-parity (LHT) decays to either Standard Model (SM) gauge boson pairs, or to
SM fermions via loop diagrams. We make a detailed study of the collider
signatures where the $A_H$ can be reconstructed from invariant mass peaks in
the opposite sign same flavor dilepton or the four-lepton channel. This
enables us to obtain information about the fundamental symmetry breaking scale
$f$ in the LHT and thereby the low-lying mass spectrum of the theory. In
addition, indication of the presence of the WZW term gives us hints of the
possible UV completion of the LHT via strong dynamics. The crucial observation
is that the sum of all production processes of heavy T-odd quark pairs has a
sizeable cross-section at the LHC and these T-odd particles eventually all
cascade decay down to the heavy photon $A_H$.  We show that for certain
regions of the parameter space with either a small $f$ of around $500 \gev$  or relatively light T-odd quarks with a mass of around $400 \gev$, one can reconstruct the $A_H$ even at the early LHC run with $\sqrt{s}=10 \tev$ and a modest integrated luminosity of $200\pb^{-1}$. At $\sqrt{s} = 14\tev$ and with an integrated luminosity of $30 \fb^{-1}$, it is possible to cover a large part of the typical parameter space of the LHT, with the scale $f$ up to $1.5\tev$  and with T-odd quark masses almost up to $1\tev$. In this region of the parameter space, the mass of the reconstructed $A_H$ ranges from $66 \gev$ to $230 \gev$.
\end{abstract}

\pagestyle{plain}

\end{titlepage}


\setcounter{page}{1}

\renewcommand{\thefootnote}{\arabic{footnote}}
\setcounter{footnote}{0}

\section{Introduction}
\label{sec:intro}

Little Higgs models~\cite{LH_original, LH_reviews} have been proposed a few
years ago to explain electroweak symmetry breaking and, in particular, to
solve the so-called little hierarchy problem~\cite{LEP_paradox}. We can view
the Standard Model (SM) as an effective field theory (EFT) with a cutoff
$\Lambda$ and parametrize new physics in terms of higher-dimensional operators
which are suppressed by inverse powers of $\Lambda$. Precision tests of the SM
have not shown any significant deviations, which in turn translates into a
cutoff of about $\Lambda \sim 5-10\tev$ which is more than an order of
magnitude above the electroweak scale. Since radiative corrections to the
Higgs mass are quadratically sensitive to the cutoff $\Lambda$, some amount of
fine-tuning is needed to get a Higgs which is lighter than about $200\gev$ as
indicated by electroweak precision data.

Little Higgs models suggest a way of stabilizing the mass of the Higgs in the
presence of a cut-off $\Lambda$ of the above kind of magnitude. Here the Higgs
particle is a pseudo-Goldstone boson of a global symmetry $G$ which is
spontaneously broken at a scale $f$ to a subgroup $H$. This symmetry protects
the Higgs mass from getting quadratic divergences at one loop. The electroweak
symmetry is broken via the Coleman-Weinberg mechanism~\cite{Coleman_Weinberg}
and the Higgs mass is generated radiatively, which leads naturally to a light
Higgs boson $m_H \sim (g^2 / 4\pi) f \approx 100\gev$, if the scale $f \sim
1\tev$. The little Higgs model can then be interpreted as an EFT up to a new
cutoff scale of $\Lambda \sim 4 \pi f \sim 10\tev$.

Among the different versions of this approach, the littlest Higgs
model~\cite{Littlest_Higgs} achieves the cancellation of quadratic divergences
with a minimal number of new degrees of freedom.  However, precision
electroweak constraints imply that the mass scale of the new particles in such
theories has to be of the order of $f \gapprox 5\tev$ in most of the natural
parameter space~\cite{LH_EW_tests}, thus necessitating fine-tuning once
more. The problem is circumvented through the introduction of an additional
discrete symmetry, the so-called T-parity~\cite{T_parity,LHT_Low}, whereby all
particles in the spectrum are classified as T-even or T-odd. This allows one
to have the Higgs mass protected from quadratic divergences, and at the same
time see a spectrum of additional gauge bosons, scalars and fermions, in the
mass range of a few hundred GeV, with the lightest T-odd particle (LTP),
typically the heavy neutral partner of the photon, $A_H$, being stable. In
particular, for the Littlest Higgs model with T-parity
(LHT)~\cite{LHT_Low,Hubisz_Meade,Hubisz_et_al,Chen_Tobe_Yuan} it was shown in
Refs.~\cite{Hubisz_et_al,Asano_et_al,Hundi_Mukhopadhyaya_Nyffeler}, that a
scale $f$ as low as $500\gev$ is compatible with electroweak precision
data. Furthermore, if T-parity is exact, the LTP can also be a potential dark
matter (DM) candidate~\cite{Hubisz_Meade,Asano_et_al,LH_Dark_Matter}.

The experimental signals at colliders of a scenario with exact T-parity have
close resemblance with those of supersymmetry (SUSY) with conserved R-parity
or universal extra dimensions (UED) with KK-parity. First of all, T-odd
particles can only be produced in pairs. Furthermore, all T-odd particles
cascade decay down to the LTP which then carries away substantial missing
transverse momentum, accompanied by jets and / or leptons rendered hard through
the release of energy in the decay of the heavy new particles.

It was later pointed out in Ref.~\cite{Hill_Hill} that T-parity can be
violated in the EFT by topological effects related to anomalies in the
underlying theory (UV completion of the LHT). This induces a
Wess-Zumino-Witten (WZW) term~\cite{WZW} in the low-energy effective
Lagrangian, similarly to the WZW term of odd intrinsic parity in the usual
chiral Lagrangian for QCD. This term encodes the Adler-Bell-Jackiw 
chiral anomaly~\cite{ABJanomaly} within the EFT framework and describes, for
instance, the decay $\pi^0 \to \gamma\gamma$. The structure of the WZW term is
thereby uniquely determined by the symmetry breaking pattern $G \to H$ and the
gauged subgroups, up to a multiplicative quantized constant. This constant is
related to the representation of new fermions in the underlying theory, if we
assume that it is strongly interacting and a fermion condensate forms which
signals spontaneous symmetry breaking. Essentially, as in QCD, the
prefactor of the WZW term is a function of the number of `colors' in the
underlying theory.

Of course, there is no such WZW term, if there are no chiral anomalies in the
underlying theory, for instance, if the low-energy non-linear sigma model
Lagrangian of the LHT derives from a linear sigma model with new heavy
fundamental scalar fields which break the symmetry, see
Ref.~\cite{LHT_no_T-violation} for an explicit construction of such an UV
completion with unbroken T-parity.

The phenomenology at colliders of the LHT with T-parity violation changes
completely~\cite{Hill_Hill, Barger_Keung_Gao, Zprime_to_ZZ,
  Freitas_Schwaller_Wyler}. Assuming that $A_H$ is the LTP, the T-violating
terms will lead to its decay into SM particles either directly into two
electroweak gauge bosons or via one-loop graphs into SM fermion pairs. The
decay width will be very small, of the order of eV. This is due to the small
prefactor in front of the WZW term which counts as order $p^4$ in the chiral
expansion, i.e.\ it is of the same size as one-loop effects in the
EFT. Nevertheless, $A_H$ will promptly decay inside the detector and one does
therefore not expect events with large missing transverse energy. As we will
see, this allows one to reconstruct the masses of the new particles, in
particular of $A_H$ itself.  On the other hand, since the T-violating
couplings are very small, the production mechanism of T-odd particles is
essentially unchanged from the case with exact T-parity and the T-odd
particles will again cascade decay down to $A_H$. Of course, with T-parity
violation, the unstable LTP will now no longer be a suitable DM candidate.

The phenomenology of the LHT with exact T-parity at the Large Hadron Collider
(LHC) has been studied quite extensively~\cite{Freitas_Wyler, Belyaev_et_al,
LHT_at_LHC, recent_LHT}. Efforts are also on to discriminate the LHT signals
from those of other scenarios where large missing $E_T$ is
predicted~\cite{LHT_vs_SUSY_UED}.  However, relatively few studies have taken
place on the collider signals of the scenario with T-parity
violation~\cite{Barger_Keung_Gao, Zprime_to_ZZ,
Freitas_Schwaller_Wyler}. Although Ref.~\cite{Freitas_Schwaller_Wyler} gives a
very comprehensive list of signals for several regions in the parameter space,
definitely more detailed studies of this interesting possibility of New
Physics are required, in particular, since the WZW term is a direct window
into the UV completion of the LHT.

In this paper we study the decay of $A_H$ into a pair of light leptons
(electrons, muons) or, via its decay into a pair of $Z$-bosons, eventually
into four leptons. With appropriate cuts, in particular demanding a large
effective mass $M_{eff} > 1\tev$, to reduce the SM background from $t\bar{t}$
for dileptons and $ZZ$ for four leptons, a clear peak in the dilepton or
four-lepton invariant mass distribution emerges at $M_{A_H}$. Since $M_{A_H}
\sim f$, this will therefore allow one to directly determine the symmetry
breaking scale $f$ of the LHT with good precision. The crucial point of our
analysis is the observation that since pairs of T-odd heavy quarks can be
produced in strong interaction processes and since all of them eventually
decay into a pair of $A_H$ bosons and various SM particles, we get a sizeable
signal at the LHC running at $10\tev$ or $14\tev$, if we add all production
processes of heavy T-odd quark pairs and look at the inclusive signal of $2l +
X$ or $4l + X$. For the decay chain $A_H \to ZZ \to 4l$, the idea to look at
the sum of all processes of T-odd heavy quark production was already proposed
in Ref.~\cite{Zprime_to_ZZ}. Based on some rough event and detector
simulations it was argued there that a signal can be seen at the LHC, in
particular, if a tight cut is applied that the four-lepton invariant mass
should lie in the narrow window $M_{A_H} \pm 6\gev$ in order to reduce the SM
background from $ZZ$. Our study will be much more detailed and we will also
include the dilepton signal which will be important for $A_H$ masses below
about $120\gev$. In this respect our analysis also differs from
Ref.~\cite{Freitas_Schwaller_Wyler} which looked into specific final states
with multiple leptons and jets coming from various decay chains, but not at
the total of all production processes leading to $A_H$.

The important points that are emphasized in this work, and where we have
gone beyond the earlier studies, are as follows: 

\begin{itemize}
\item We have performed a detailed study of two-and four-lepton signals
for the situation where the LTP (the heavy photon $A_H$) is liable to decay.
The relevant backgrounds and ways of reducing them have also been investigated.

\item Our simulation shows how the mass of the $A_H$ can be reconstructed from
peaks in the dilepton and four-lepton spectra in different regions of the
parameter space. This allows us to determine the fundamental
parameter $f$ in the LHT and thus gain deeper insight into the model 
from its low-lying particle spectrum.

\item The clear identification of a decaying $A_{H}$ suggests the breaking 
of T-parity via the WZW terms, and thus a possible UV completion of the theory 
in the form of a strong dynamics at a higher scale.

\item The confirmation of a decaying LTP sets the scenario clearly apart from
R-parity conserving SUSY or UED with conserved KK-parity. Moreover, the
observation of invariant mass peaks in dilepton and four-lepton channels is
not expected in R-parity violating SUSY either.
\end{itemize}

This paper is organized as follows. In Sec.~\ref{sec:overview_LHT} we briefly
review some basics about the Littlest Higgs model with exact T-parity and then
sketch how T-parity is violated by the WZW anomaly term, leading to an
unstable $A_H$.  Section~\ref{sec:2l_4l_processes} discusses the dilepton and
four-lepton signal processes, starting with the parton-level production of
heavy T-odd quark pairs and the decay modes and branching ratios of $A_H$. We
also argue why the $A_H$ is different from the often considered $Z^\prime$
gauge boson or the first Kaluza-Klein excitation of the graviton and thus
could have escaped detection, even with a low mass of the order of
$100\gev$. Finally, we present our choice of benchmark points for several
values for the heavy quark masses $m_{q_H}$ and for several values for the
mass of the heavy photon $M_{A_H}$. Section~\ref{sec:event_generation} gives
details on our event generation for the signal and the background. We describe
the main sources of backgrounds from the SM and from within the LHT. We then
go on to present our event selection criteria for the dilepton and the
four-lepton signal and the various cuts to reduce the backgrounds. In
Sec.~\ref{sec:results} we present our results, first for the dilepton signal
and then for the four-lepton signature. In both cases, we give numbers for the
expected signal and background cross-sections after the cuts for the LHC
running at a center of mass energy of $10\tev$ ($14\tev$) and the number of
signal and background events for an integrated luminosity of $200\pb^{-1}$
($30\fb^{-1}$).  In particular, we will show that in the case where the heavy
T-odd quarks are not much above the bound of approximately $m_{q_H} > 350\gev$
from Tevatron, as estimated in Ref.~\cite{Freitas_Schwaller_Wyler}, even with
a rather modest luminosity of $200\pb^{-1}$, one will get a signal in the
early run of the LHC at $10\tev$. On the other hand, for T-odd quarks with
masses around $1000 \gev$, a clear signal will be visible with $30 \fb^{-1}$
of integrated luminosity for the LHC running at $14\tev$.  We summarize and
conclude in Section~\ref{sec:conclusions}.


\section{The Littlest Higgs model with T-parity and T-parity violation} 
\label{sec:overview_LHT}

\subsection{The Littlest Higgs model with T-parity}

In the LHT a global symmetry $SU(5)$ is spontaneously broken down to $SO(5)$
at a scale $f \sim 1\tev$. An $[SU(2) \times U(1)]^2$ gauge symmetry is
imposed, which is simultaneously broken to the diagonal subgroup $SU(2)_L
\times U(1)_Y$, the latter being identified with the SM gauge group. This leads
to four heavy gauge bosons $W_H^\pm, Z_H$ and $A_H$ with masses $\sim f$ in
addition to the SM gauge fields. The SM Higgs doublet is part of an assortment
of pseudo-Goldstone bosons which result from the spontaneous breaking of the
global symmetry. This symmetry protects the Higgs mass from getting quadratic
divergences at one loop, even in the presence of gauge and Yukawa
interactions. The multiplet of Goldstone bosons contains a heavy $SU(2)$
triplet scalar $\Phi$ as well. In contrast to SUSY, the new states which
cancel the quadratically divergent contributions to the Higgs mass due to the
top quark, gauge boson and Higgs boson loops, respectively, are heavy
fermions, additional gauge bosons and triplet Higgs states.

In order to comply with strong constraints from electroweak precision data on
the Littlest Higgs model~\cite{LH_EW_tests}, one imposes
T-parity~\cite{T_parity} which maps the two pairs of gauge groups $SU(2)_i
\times U(1)_i, i=1,2$ into each other, forcing the corresponding gauge
couplings to be equal, with $g_1 = g_2$ and $g_1^\prime = g_2^\prime$. All SM
particles, including the Higgs doublet, are even under T-parity, whereas the
four additional heavy gauge bosons and the Higgs triplet are T-odd. The top
quark has two heavy fermionic partners, $T_{+}$ (T-even) and $T_{-}$
(T-odd). For consistency of the model, one has to introduce the additional
heavy, T-odd vector-like fermions $u^i_H, d^i_H, e^i_H$ and $\nu^i_H$
($i=1,2,3$) for each SM quark and lepton field. For further details on the
LHT, we refer the reader to 
Refs.~\cite{LHT_Low,Hubisz_Meade,Hubisz_et_al,Chen_Tobe_Yuan}. 

The masses of the heavy gauge bosons in the LHT are given by 
\be \label{gauge_boson_masses}
M_{W_H} = M_{Z_H} = g f \left(1 - {v^2 \over 8 f^2} \right) \approx 0.65 f,
\qquad  
M_{A_H} = {f g^\prime \over \sqrt{5}} \left(1 - {5 v^2 \over 8 f^2} \right)  
\approx 0.16 f, 
\ee
where corrections of ${\cal O}(v^2/f^2)$ are neglected in the approximate
numerical values. Thus these particles have masses of several hundreds of GeV
for $f \sim 1\tev$, although $A_H$, the heavy partner of the photon,
can be quite light, because of the small prefactor, and is usually assumed to
be the LTP. The masses of the heavy, T-odd fermions are determined by general
$3\times 3$ mass matrices in the (mirror) flavor space, $m_{q_H,l_H}^{ij}
\sim \kappa_{q,l}^{ij} f$ with $i,j=1,2,3$. We simplify our analysis by
assuming that $\kappa_q^{ij} = \kappa_q \delta^{ij}$. The parameter $\kappa_q
\sim {\cal O}(1)$ thus determines the masses of the heavy quarks in the
following way: 
\be \label{mirror_quark_masses}
m_{u_H} = \sqrt{2} \kappa_q f \left(1 - {v^2 \over 8 f^2} \right), 
\qquad 
m_{d_H} = \sqrt{2} \kappa_q f, 
\ee
thereby allowing the new heavy quarks to have masses ranging from several
hundreds of GeV to a TeV, for $f \sim 1\tev$.  Similarly, the masses of the
heavy leptons in the spectrum are determined by a common parameter
$\kappa_l$. Note that these heavy quarks and leptons cannot be decoupled from
the model as there is an upper bound $\kappa \leq 4.8$ (for $f = 1\tev$)
obtained from 4-fermion operators~\cite{Hubisz_et_al}. We will come back to
lower limits on the masses of the heavy quarks and therefore on $\kappa_{q}$
in the context of the model with T-parity violation. 

While they can act as a source of model background for our leptonic signals,
the T-odd leptons do not otherwise play any important role in our analysis.
We will therefore use $\kappa_l = 1$ throughout. In the section on background
analysis, we discuss the model backgrounds in further detail. Thus $f$ and
$\kappa_q$ determine the part of the LHT spectrum relevant for our study. The
mass of the triplet scalar $\Phi$ is related to the doublet Higgs mass by
$m_\Phi = \sqrt{2} m_H f / v$.  We will take $m_H = 120\gev$ throughout this
paper. Two more dimensionless parameters $\lambda_1$ and $\lambda_2$ appear in
the top quark sector; the top mass being given by $m_t = (\lambda_1 / \sqrt{1
+ R^2}) v$, where $R = \lambda_1 / \lambda_2$. The masses of the two heavy
partners of the top quark, $T_{+}$ and $T_{-}$, can be expressed as $m_{T_{+}}
= \lambda_2 \sqrt{1 + R^2} f$ and $ m_{T_{-}} = \lambda_2 f$.  We use $m_t =
175\gev$ in our analysis and set $R = 1$.

\subsection{T-parity violation}
\label{sec:T_violation}

T-parity violation in the LHT and thus the decay of the heavy photon $A_H$
arises via the so-called Wess-Zumino-Witten term~\cite{WZW}, which, according
to Ref.~\cite{Hill_Hill}, can be written as follows:
\be \label{WZW_term}
\Gamma_{\rm WZW} = {N \over 48 \pi^2} \left( \Gamma_0[\Sigma] + \Gamma[\Sigma,
A_l, A_r] \right). 
\ee
The functional $\Gamma_0[\Sigma]$ is the ungauged WZW term which depends only
on the non-linear sigma model field $\Sigma$. It cannot be expressed as a
four-dimensional integral over a local Lagrangian. Instead, a closed form can
be written as an integral over a five-dimensional manifold with ordinary
spacetime as its boundary~\cite{WZW}. The term $\Gamma[\Sigma, A_l, A_r]$ is
the gauged part of the WZW term. This part can be written as an ordinary
four-dimensional integral over a local Lagrangian.  The explicit expressions
for the functionals and the relation of the fields $A_{l,r}$ to the gauge
fields in the LHT can be found in Ref.~\cite{Hill_Hill}. While these
functionals are uniquely given by the symmetry breaking pattern $SU(5) \to
SO(5)$ and the gauged subgroups in the LHT, the integer  $N$ in
Eq.~(\ref{WZW_term}) depends on the UV completion of the LHT. In strongly
coupled underlying theories it will be related to the representation of the
fermions whose condensate acts as order parameter of the spontaneous symmetry
breaking. In the simplest case, $N$ will simply be the number of `colors' in
that UV completion, as is the case for the WZW term in ordinary QCD. The
overall coefficient $N / 48 \pi^2$ encapsulates the effect of the
chiral anomaly, which is a one-loop effect in the corresponding 
high-scale theory.

As noted in Ref.~\cite{Hill_Hill}, the WZW term in Eq.~(\ref{WZW_term}) is not
manifestly gauge invariant. Gauge invariance is violated by terms with three
or four gauge bosons with an odd number of T-odd gauge bosons, e.g.\ by a term
like $\epsilon_{\mu\nu\rho\sigma} V_H^\mu V^\nu \partial^\rho V^\sigma$, where
$V_H$ is a T-odd gauge boson and $V$ denotes a SM gauge boson. Such anomalous
terms need to be cancelled to have a consistent theory and some mechanisms to
achieve this are discussed in Ref.~\cite{Hill_Hill}. After this cancellation,
the leading T-odd interactions appear only at order $1/f^2$. For instance, we
get a vertex with one T-odd gauge boson and two SM gauge bosons from
$\epsilon_{\mu\nu\rho\sigma} (H^\dagger H / f^2) V_H^\mu V^\nu \partial^\rho
V^\sigma$, after the Higgs doublet $H$ gets a vacuum expectation value $v$.

To leading order in $1/f$, the part of the WZW term containing one neutral
T-odd gauge boson is given, in unitary gauge, by 
\bea
\lefteqn{\Gamma_n  =  {N g^2 g^\prime \over 48 \pi^2 f^2} \int d^4x \, (v +
  h)^2 \, \epsilon_{\mu\nu\rho\sigma} \times \nonumber} \\
& & \left[ -(6/5) A_H^\mu \left( c_w^{-2} Z^\nu \partial^\rho Z^\sigma +
  W^{+\nu} D_A^\rho W^{-\sigma} + W^{-\nu} D_A^\rho W^{+\sigma} + i(3 g x_w +
  g^\prime s_w) W^{+\nu} W^{-\rho} Z^\sigma \right) \right. \nonumber \\ 
& & \left. + t_w^{-1} Z_H^\mu \left( 2 c_w^{-2} Z^\nu \partial^\rho Z^\sigma +
  W^{+\nu} D_A^\rho W^{-\sigma} + W^{-\nu} D_A^\rho W^{+\sigma} - 2i (2gc_w +
  g^\prime s_w) W^{+\nu} W^{-\rho} Z^\sigma \right) \right]. \nonumber \\
& & 
\label{WZW_A_H}
\eea
Here $h$ is the physical Higgs boson, $D_A^\mu W^{\pm\nu} = (\partial^\mu \mp
i e A^\mu)W^{\pm\nu}$ and $s_w, c_w$ and $t_w$ denote the sine, cosine and
tangent of the weak mixing angle, respectively.  All T-violating vertices with
up to four legs have been tabulated in Ref.~\cite{Freitas_Schwaller_Wyler} and
implemented into a model file for CalcHEP 2.5~\cite{CalcHEP,
CalcHEP_T_violation}.

If $A_H$ is heavy, the vertices in Eq.~(\ref{WZW_A_H}) lead to its decay into
a pair of $Z$-bosons or into $W^+ W^-$ with a decay width of the order of
eV~\cite{Barger_Keung_Gao}. On the other hand, if $M_{A_H} < 2 M_W$, the heavy
photon cannot decay into on-shell SM gauge bosons. It could still decay into
(one or two) off-shell SM gauge bosons, but for low masses loop induced decays
into SM fermions will dominate. In fact, as discussed in
Ref.~\cite{Freitas_Schwaller_Wyler}, the T-violating vertices can couple the
$A_H$ to two SM fermions via a triangle loop. But since the corresponding
one-loop diagrams are logarithmically divergent, one needs to add counterterms
to the effective Lagrangian of the form
\bea
{\cal L}_{\rm ct} & = & \bar{f} \gamma_\mu \left( c_L^f P_L + c_R^f P_R
\right) f A_H^\mu, \label{counterterms} \\ 
c_i^f & = & c_{i,\epsilon}^f \left( {1 \over \epsilon} + \log(\mu^2) +
\order{1} \right), \label{coeff_CT}
\eea
where $P_{L,R} = (\mathbf{1} \mp \gamma_5)/2$.  As shown in
Ref.~\cite{Freitas_Schwaller_Wyler} the counterterms can also be written in a
manifestly gauge invariant way. These counterterms are only some of the
infinitely many terms which have to be included anyway at higher orders in the
momentum and loop expansion in the EFT. This procedure to renormalize the EFT
order by order is well known from chiral perturbation theory~\cite{ChPT} and,
as usually done there, dimensional regularization was used in
Ref.~\cite{Freitas_Schwaller_Wyler} which preserves chiral and gauge
invariance.\footnote{Of course, in view of the notorious problems with
dimensional regularization in chiral gauge theories and the appearance of
$\gamma_5$ and the Levi-Civita tensor $\epsilon_{\mu\nu\rho\sigma}$, for a
consistent treatment of divergent loop integrals a more appropriate 
regularization scheme should be chosen like the proper-time method or 
zeta-function regularization, see Ref.~\cite{other_regularizations}.}

The coefficients $c_i^f(\mu)$ of the counterterms can be estimated by naive
 dimensional analysis~\cite{NDA} or naturalness arguments. Since the scale
 dependence of the loop diagrams is cancelled by the scale dependence of the
 counterterms $c_i^f(\mu)$, any change of order one in the renormalization
 scale should be compensated by a change of order one in
 $c_i^f(\mu)$. Therefore these coefficients are given, up to $\order{1}$
 factors, by the coefficients of the leading $1/\epsilon$ divergence in
 dimensional regularization of the loop integrals. The coefficients
 $c_{i,\epsilon}^f$ are explicitly listed in
 Ref.~\cite{Freitas_Schwaller_Wyler} and we have included the vertices from
 Eq.~(\ref{counterterms}) in the CalcHEP model file. 

The finite parts of the counterterms are determined by the underlying
theory. For a given UV completion of the LHT, they can be obtained, in
principle, by integrating out the new `resonances' which lie above the cutoff
$\Lambda$ of the LHT. Since the branching ratios (BR's) into the different SM
fermions depend on these coefficients, at least in principle, we could get
information on the UV completion of the LHT by precisely measuring the BR's of
$A_H$. We will discuss below, how $\order{1}$ changes (different for quarks
and leptons) in the coefficients $c_{i,\epsilon}^f$ could affect these BR's
and thus our analysis. Note that the unknown constant $N$ from
Eq.~(\ref{WZW_A_H}) cancels in the branching ratios. Actually, if we could
measure the total decay width of $A_H$, we could even get information on $N$
itself, in the same way as the decay $\pi^0 \to \gamma\gamma$ yields
information about the number of colors in ordinary QCD, however, the width of
$\order{\mbox{eV}}$ for $A_H$ is too small to be measurable.

The prefactor of the WZW term, $N/48\pi^2$, is of the size of a one-loop
effect, thus the coupling of $A_H$ and other T-odd gauge bosons to SM fermions
via a triangle-loop is effectively 2-loop suppressed. Therefore these
T-violating couplings will not affect the production mechanism of T-odd
particles and their cascade decays at colliders, or EW precision
observables~\cite{Freitas_Schwaller_Wyler}. In particular, T-parity violation
should still satisfy the EW data with a rather small scale~$f$.  It is only in
decays of the $A_H$ that the anomaly term acquires phenomenological
importance.  As we demonstrate in what follows, reconstruction of the $A_H$
mass becomes possible through such decays, thus confirming the bosonic nature
of the LTP.

Since there is no stable LTP now, the collider signals in the LHT with
T-parity violation are completely different from the LHT with exact
T-parity. The LEP bounds of order $100\gev$ still apply to all T-odd particles
except for the $A_H$. In addition, based on an analysis of recent
Tevatron data from CDF Vista on multijet events~\cite{CDF_Vista}, it has been
argued in Ref.~\cite{Freitas_Schwaller_Wyler} that a bound of $m_{q_H} >
350\gev$ applies to the LHT with broken T-parity.


\section{The dilepton and four-lepton signal processes}
\label{sec:2l_4l_processes}

\subsection{Parton level production of heavy T-odd quark pairs}

As the cross-section for direct single or pair production of $A_H$ is very
tiny, this gauge boson can essentially only be produced via the decay of
heavier T-odd particles.  Hence, in principle, we should be considering the
production of all such T-odd particle pairs and their subsequent decays. But
owing to the substantial technical difficulties in simulating all such
processes together, we restrict our attention to the production of heavy T-odd
quarks in the initial parton level hard scattering. Needless to say, our
cross-sections for the specific final states that we consider are then rather
underestimated, and can be taken as lower bounds.

We consider the following processes for the production of T-odd quark pairs at
the LHC: 
\bea \label{qHbarqH} 
pp & \rightarrow & q_{H}\bar{q}_H + X, \qquad \mbox{where} \ \  
q_{H}=u_{H},d_{H},c_{H},s_{H},b_{H},t_{H},   \\ 
pp & \rightarrow & u_H u_H + X, \ \ \bar{u}_H \bar{u}_H + X, \ \ d_H d_H + X,
\ \ \bar{d}_H \bar{d}_H + X,  \nonumber \\
& & u_H d_H + X, \ \ \bar{u}_H \bar{d}_H + X, \ \ u_H \bar{d}_H + X, 
\label{other_processes}
\eea
where $t_H$ denotes the lighter T-odd partner of the top quark.  Since,
$T_{-}$, the heavier T-odd partner of the top quark has a mass of 1013 GeV
(for $f = 1\tev$) and is thus heavier than $t_H$ (for most of our choices for
$\kappa_q$ below), its cross-section is much smaller and we have neglected its
pair production. Of course, for lower values of $f$, both $T_{-} \bar{T}_{-}$
and heavy T-odd gauge boson productions can have appreciable cross-section at
the LHC.

In general, we expect the processes in Eq.~(\ref{qHbarqH}) to be dominant
because of the strong interaction production channels through gluon-gluon
fusion and $q\bar{q}$-annihilation. But the electroweak processes from
Eq.~(\ref{other_processes}) also contribute significantly to the cross-section
via $t$-channel T-odd gauge boson exchange, especially when the T-odd gauge
bosons become relatively light, i.e.\ for low values of $f$. For instance, for
$f= 1\tev$ and $\kappa_q=0.5$, we find $m_{q_H} \sim 700\gev$ and the total
production cross-section for a pair of heavy T-odd quarks is about
$2.1~(0.7)\pb$ for the LHC running at $14~(10)\tev$. On the other hand, we can
also obtain $m_{q_H} \sim 700\gev$ by choosing $f=500\gev$ and
$\kappa_q=1$. In this later case the cross-section goes up to $5.8~(2.3)\pb$,
where actually the electroweak processes from Eq.~(\ref{other_processes}) are
found to dominate. This fact has also been observed recently in
Ref.~\cite{recent_LHT}. We will come back to this point below. If the heavy
T-odd quarks do not lie much above the lower bound of $m_{q_H} > 350\gev$, we
get a cross-section of $36~(13)\pb$ for $m_{q_H} \sim 400\gev$ with
$f=1\tev$. For details on the production of the heavy quarks and the relevant
plots for the variation of these cross-sections with $f$ or $m_{q_H}$, we
refer the reader to the Refs.~\cite{Freitas_Wyler, Belyaev_et_al}. We have
checked that for individual processes, our results agree with them. Note,
however, as $f$ determines the mass of $A_H$, and in turn its decay modes, in
order to study the effect of variation of quark masses with $M_{A_H}$ fixed,
in some cases we vary $\kappa_q$ to adjust the masses of the heavy T-odd
quarks.

The important fact is that the sum of the cross-sections of all T-odd
quark pair production processes can be sizeable, in particular for not
too large $m_{q_H}$. This will allow us to extract a clear signal
after cuts are applied. Furthermore, since additional electroweak
processes which can, for instance, produce pairs of T-odd gauge bosons
$V_H V_H$, finally also lead to two $A_H$'s, the given cross-sections,
as mentioned before, are actually rather lower bounds.

The initially produced T-odd heavy quarks subsequently decay as $q_H \to W_H
q^\prime, Z_H q, A_H q$ and then $W_H \to A_H W$, $\bar {q} q_{H}^{\prime}$
and $Z_H \to A_H h$, $q \bar{q}_{H}$.  At one point in such decay chains of a
pair of $q_H$'s, we are left with two $A_H$ bosons, which will further decay
as discussed in the next subsection. There will also be several hard jets and
leptons and some amount of missing $E_T$, if there are decays of $W^\pm$ and
$Z$ into neutrinos.

The initial parton level hard-scattering matrix elements and the relevant
decay branching ratios for the signal in the LHT with T-parity violation are
calculated with the help of CalcHEP (Version 2.5.1)~\cite{CalcHEP}. We have
used the CalcHEP model files for the LHT (with exact T-parity) from
Ref.~\cite{Belyaev_et_al}\footnote{A new CalcHEP model file has been written
by the authors of Ref.~\cite{recent_LHT} which includes some missing factors
of order $v^2/f^2$ in the couplings of T-odd fermions to the $Z$- and
$W$-bosons, which were found in Ref.~\cite{correction_Feynmanrules}. These
changes in the Feynman rules will, however, not significantly affect our
analysis, which focuses on the decays of the $A_H$ boson.} and the one from
Refs.~\cite{Freitas_Schwaller_Wyler,CalcHEP_T_violation} for the T-violating
terms. We have used the leading order CTEQ6L~\cite{CTEQ} parton distribution
functions with NLO running of $\alpha_{s}$ with $\alpha_s(M_Z) = 0.118$. The
QCD factorization and renormalization scales were set equal to the sum of the
masses of the particles which are produced in the initial parton level
scattering process.

\subsection{Decay modes of $A_H$}

A comprehensive list of possible final states in the LHT with T-parity
violation after the decay of the $A_H$'s is given in
Ref.~\cite{Freitas_Schwaller_Wyler}. Here we are interested in either dilepton
or four-lepton signals from the decay of the two $A_H$'s.  The advantage of
these leptonic decay channels of $A_{H}$ is very apparent. As we will see, one
can obtain clean signatures over the backgrounds with a minimal number of
selection cuts and with luminosity building up, clear peaks in the
invariant-mass distributions of dileptons or four leptons give us information
about the $A_{H}$ mass and thus on the symmetry breaking scale $f$.  The decay
branching fractions of $A_H$ to either leptons (electron, muon) or to a pair
of $Z$ bosons (one of which might be off-shell) are given in
Table~\ref{tab:A_H-Decay}. Note that the BR for the further decay $ZZ \to l^+
l^- {l^\prime}^+ {l^\prime}^-$, where $l, l^\prime=\{e,\mu\}$, is only $4.5
\times 10^{-3}$, but the signal in this channel is very clean and the SM
backgrounds, primarily from $ZZ$ production, can be reduced efficiently as we
will discuss below.

\begin{table}[htb]
\begin{center}
\begin{tabular}{|c|c|c|r@{.}l|}
\hline
 f& $M_{A_{H}}$&$\BR(A_H \to e^+ e^-) + \BR(A_H \to \mu^+ \mu^-)$ &
 \multicolumn{2}{c|}{$\BR(A_{H} \to ZZ^{(*)})$ } \\  
(GeV) &(GeV)& (\%) &\multicolumn{2}{c|}{(\%)} \\
\hline 
500 &66& 7.59 & \multicolumn{2}{c|}{$\sim 0$}  \\
\hline
750 & 109& 7.40 & \hspace*{1.2cm} 0&18  \\
\hline 
1000 &150& 3.42 & 11&03  \\
\hline
1100 &166& 0.99 & 8&67  \\
\hline 
1500 & 230& 0.02 & 22&45 \\
\hline 
\end{tabular}
\caption{Decay branching fractions of $A_{H}$ to leptons ($l=e,\mu$) or
  $ZZ^{(*)}$ as a function of the scale $f$, i.e.\ the mass $M_{A_H}$.} 
\label{tab:A_H-Decay}
\end{center}
\end{table}

As already observed in Ref.~\cite{Freitas_Schwaller_Wyler}, for lower masses
($M_{A_H} \lapprox 120 \gev, f \lapprox 800 \gev$) the decay of $A_{H}$ is
dominated by the loop-induced two-body modes into fermions, whereas for higher
masses ($M_{A_H} > 2M_{W} , f > 1070 \gev$) the two-body modes to gauge boson
pairs dominate.  For intermediate masses, both the two-body and three-body
modes compete (in the three-body mode we have one on-shell $W^{\pm}$ or
$Z$). The decay into two off-shell $Z$'s for low $f$ will have a very small
branching-fraction, as the relative one-loop suppression is already
compensated by the off-shellness of one vector boson.

As mentioned earlier, the decay rates of $A_H$ into fermions (quarks, charged
leptons, neutrinos) via one-loop triangle diagrams depend on the values of the
finite terms in the counterterms from Eq.~(\ref{counterterms}). To obtain the
results given in Table~\ref{tab:A_H-Decay} we followed
Ref.~\cite{Freitas_Schwaller_Wyler} and used naive dimensional analysis to fix
the $\order{1}$ constants from Eq.~(\ref{coeff_CT}) to be exactly equal to
one.  These finite terms are determined by the UV completion of the LHT and
could easily be different from one. In particular, one could imagine a
situation, where the underlying theory couples differently to quarks and
leptons. For instance, it could happen that all the couplings of the charged
leptons could be bigger by a factor of two, which is well within the
uncertainty of naive dimensional analysis. This would increase the partial
decay width $\Gamma(A_H \to \mbox{all charged leptons})$ by a factor of
four. The corresponding change of the branching ratio $\BR(A_H \to e^+ e^-) +
\BR(A_H \to \mu^+ \mu^-)$, which is relevant for our study, depends on the
total decay width and therefore on the mass of $A_H$ or the scale $f$. It
increases by about a factor of three for $M_{A_H} = M_Z$, i.e.\ for
$f=650\gev$. Such a scenario would of course require less luminosity to get a
certain number of dilepton events in our analysis below. On the other hand, if
the underlying theory increases the couplings of $A_H$ to only the quarks by a
factor of two compared to naive dimensional analysis, then the $\BR(A_H \to
e^+ e^-) + \BR(A_H \to \mu^+ \mu^-)$ would be smaller (by a factor of about
three for $M_{A_H} = M_Z$) and we would need more luminosity. While the
precise numbers for these fermionic branching ratios therefore crucially
depend on the unknown $\order{1}$ coefficients in the counterterms, the
overall results of our analysis are not expected to change. In particular, the
required luminosity is not expected to change by more than a factor of three,
up or down.

As already noted in Ref.~\cite{Freitas_Schwaller_Wyler} as soon as the
$WW^{(*)}$ and $ZZ^{(*)}$ decay channels for $A_H$ open up for larger
$M_{A_H}$ or $f$, they quickly dominate over the fermionic modes. Therefore
the overall picture and the value of $f$ where this cross-over occurs, does
not depend too sensitively on the precise values of the counterterms, as long
as they vary only in a reasonably small window around the values as predicted
by naive dimensional analysis.

\subsection{Why is the $A_H$ different from a usual $Z^\prime$ ?}
\label{sec:AH_vs_Z_prime}

Of course, the strategy to look for a resonance peak in the invariant mass
distribution of dileptons is well known from the searches for a $Z^\prime$
gauge boson which appears in many models of New Physics, see for instance the
recent reviews~\cite{Z_prime} and references therein. 

Low energy weak neutral current experiments are affected by $Z^\prime$
exchange, which is mainly sensitive to its mass, and by $Z-Z^\prime$ mixing.
On the other hand, measurements at the $Z$-pole are very sensitive to
$Z-Z^\prime$ mixing, which lowers the mass of the $Z$ relative to the SM
prediction and also modifies the $Zf\bar{f}$ vertices. For $e^+e^-$ colliders,
like LEP2, a $Z^\prime$ much heavier than the center of mass energy
would manifest itself through induced four-fermion interactions,
which then interfere with virtual $\gamma$ and $Z$ contributions for leptonic
and hadronic final states. The primary discovery mode at hadron colliders,
like the Tevatron, is from the direct Drell-Yan production of a dilepton
resonance. 

The bounds on the $Z^\prime$ mass in a variety of popular models are usually
obtained by assuming the $Z^\prime$ coupling to SM fermions to be of
electroweak strength and family universal. Then for some models the strongest
bounds come from electroweak precision tests yielding $M_{Z^\prime} \gapprox
1200-1400\gev$ at 95\% confidence level. On the other hand, for a sequential
$Z^\prime$ model, the LEP2 lower bound is even around $1800\gev$.  For other
models, the bounds from direct searches at the Tevatron are better than those
derived from electroweak data, typically one obtains $M_{Z^\prime} \gapprox
800-1000\gev$ for these models~\cite{Z_prime}. 

Why does this fact not rule out a $A_H$ with a mass around $50-250\gev$ which
we will consider below ?  The crucial point is that although  a light
$A_H$ decays with a large BR into a pair of SM fermions, the actual coupling
of $A_H$ to two SM fermions is very small. Essentially, the coupling is of the
size of a two-loop effect as discussed above. Thus the couplings are very
different from the most commonly considered $Z^\prime$ models with couplings
of electroweak strength for which the above limits apply. Therefore the direct
production cross-section of such a low-mass $A_H$ in $e^+ e^-$ colliders like
LEP2 or at the Tevatron is tiny, of the order of
$10^{-6}\pb$~\cite{Freitas_Schwaller_Wyler}. Also the four-fermion operators
induced by $Z^\prime$ at low energies have only a small coupling and will not
affect low-energy weak observables. Furthermore, the coupling of $A_H$ to $WW$
or $ZZ$ which is directly induced by the WZW term, see Eq.~(\ref{WZW_A_H}), is
very small and thus the production cross-section for $A_H$ radiated off some
$W$ or $Z$ boson is again very small. Therefore the $A_H$ cannot be produced
directly, but only via the decay of heavier T-odd particles, which themselves
have not yet been observed.

As far as the decay $A_H \to ZZ$ is concerned, again the small coupling of
$A_H$ to SM fermions leads to a tiny $s$-channel production cross-section at
LEP2 or the Tevatron. This is in contrast to the case of models with warped
extra dimensions, like Randall-Sundrum (RS)~\cite{Randall_Sundrum}, where the
first Kaluza-Klein excitation of the graviton, $G_1$, can have a sizeable
coupling to SM fermions and also often decays into $ZZ$ with a branching ratio
of typically 5\%. From the absence of any deviation from the SM signal in
$e^+ e^- \to Z Z$ at LEP2 it was concluded in Ref.~\cite{ee_ZZ_LEP2} that
$M_{G_1} > 700\gev$.  Recently, CDF~\cite{CDF_ZZ_eeee} has searched for a new
heavy particle decaying to $ZZ \to eeee$ in the mass range of
$500-1000\gev$. In $1.1 \fb^{-1}$ of integrated luminosity, no event was
observed after all the selection cuts, with an expected background of $0.028
\pm 0.014$ events. Within the RS-model, this translates into $\sigma(p\bar{p}
\to G_1) \times \BR(G_1 \to ZZ) < 4\pb$ at 95\% C.L. for $M_{G_1} \sim
500\gev$.  Since the mass region below $400\gev$ was used to control the
background from hadrons faking electrons, a potential signal from a lighter
resonance, like the $A_H$ with a mass around $150-250\gev$, could not be
observed. In any case, no signal is expected, since for a $A_H$ of mass $230
\gev$, produced in cascade decays of T-odd quarks with mass $400 \gev$,
$\sigma(p\bar{p}\to A_H) \times \BR(A_H \to ZZ)=0.03\pb$. This is well below
the above bound. A light $A_H$ with a mass well below $1\tev$, giving a 
simultaneous signal in the dilepton and four-lepton channels (via $ZZ$),
also sets the LHT with T-parity violation apart from R-parity violating  
SUSY models with an additional $Z^\prime$ which can decay into four leptons 
via a slepton/sneutrino pair, as proposed in Ref.~\cite{SUSY_Zprime_4l}.

\subsection{Choice of benchmark points}

As discussed above, the production cross-section of heavy T-odd quark pairs
decreases with increasing mass $m_{q_H}$, up to the discussed enhancement of
the electroweak processes from Eq.~(\ref{other_processes}) for low $f$. We
therefore will choose benchmark points (BP's) with $m_{q_H} \sim 400, 700,
1000\gev$ to see this effect. The point with the lightest mass is close to the
bound $m_{q_H} > 350\gev$ found in Ref.~\cite{Freitas_Schwaller_Wyler} from
recent Tevatron data.

The intermediate mass region $120 \gev < m_{A_H} < 165 \gev$ ($800 \gev < f <
1100 \gev$) will be the most difficult to analyze, since neither the BR of
$A_H$ into dileptons nor into four leptons (via $ZZ$) dominates as can be seen
from Table~\ref{tab:A_H-Decay}. Therefore we first take a benchmark point from
this region and choose $f=1\tev$ which corresponds to $M_{A_H} =
150\gev$. Later, we will also take $f=500\gev$, where the dilepton mode
dominates and $f=1500\gev$, where the dilepton mode is negligible and the
decay into $ZZ$ and thus into four leptons is important.

The first two BP's with $f=1\tev$ are chosen in order to illustrate the effect
of low and heavy quark masses both at the production level and at the level of
kinematical variables. We take for the first BP-1 $m_{q_H} = 400\gev~(\kappa_q
= 0.285)$ and for the second BP-2 $m_{q_H} = 700\gev~(\kappa_q = 0.5)$, see
Table~\ref{tab:BPs}. 

As we will see below, BP-1 with a rather low $m_{q_H}$ leads to a clean
dilepton signal over the backgrounds with rather modest luminosity. Therefore
such a scenario should be testable during the early run of the LHC with
$10\tev$ center of mass energy, even in the difficult intermediate mass region
for $M_{A_H}$. We expect the analysis to be easier for either lighter
$M_{A_H}$ (lower $f$) or heavier $M_{A_H}$ (higher $f$), where one of the
dilepton or four-lepton signals will clearly dominate.

For the further BP's, we therefore restrict ourselves to the heavier quark
masses $m_{q_H} \sim 700, 1000\gev$.  For $m_{q_H} \sim 700\gev$, we show the
possible reconstruction of the $A_{H}$ mass in the invariant mass
distributions of dileptons for a point with very low values of $f =
500\gev~(M_{A_H} = 66\gev)$ (BP-3) and of four leptons for a point with a
higher value of $f = 1500\gev~(M_{A_H} = 230\gev)$ (BP-4), see
Table~\ref{tab:BPs}. As the T-odd quarks become heavier, their production
cross-section goes down. It then becomes increasingly difficult to obtain a
reasonable number of signal events over the background. In such a scenario, in
order to check the reach of the LHC, we choose the last two benchmark points
such that $m_{q_H} \sim 1000\gev$. For reasons discussed above, here also we
consider two different values of $f$, $f = 1000\gev~(M_{A_H} = 150\gev)$
(BP-5) and $f = 1500\gev~(M_{A_H} = 230\gev)$ (BP-6).

\begin{table}[htb]
\begin{center}
\begin{tabular}{|l|r|r|r|r|r@{.}l|r@{.}l|r@{.}l|}
\hline
Benchmark &$m_{d_{H}}$& $m_{u_{H}}$
&$M_{A_{H}}$& \multicolumn{1}{c|}{$f$} &\multicolumn{2}{c|}{$\kappa_{q}$}
&\multicolumn{2}{c|}{$\sqrt{s} = 10\tev$}&
\multicolumn{2}{c|}{$\sqrt{s} = 14\tev$}\\   
Point & (GeV) &(GeV) &
(GeV)&(GeV)&\multicolumn{2}{c|}{} &\multicolumn{2}{c|}{$\sigma_{q_H q_H}$}
&\multicolumn{2}{c|}{$\sigma_{q_H q_H}$} \\
& & & & &\multicolumn{2}{c|}{} & \multicolumn{2}{c|}{(fb)} &
\multicolumn{2}{c|}{(fb)}\\ 
\hline
BP-1 & 403 & 400&  150 & 1000 &\hspace*{0.0cm} 0&285 &\hspace*{0.4cm}
12764&6&\hspace*{0.5cm}35989&0\\ 
\hline
BP-2 & 707 & 702 & 150 & 1000 &0&5   &660&8  &2061&0  \\
\hline
BP-3 & 707 & 686 & 66  & 500  &1&0   &2298&1 &5750&4 \\
\hline
BP-4 & 742 & 740 & 230 & 1500 &0&35  &373&0  &1283&1 \\
\hline
BP-5 & 1025& 1018& 150 & 1000 &0&725 &119&9  &421&0 \\
\hline 
BP-6 & 1008& 1004&230  & 1500 &0&475 &66&3  &261&0 \\ 
\hline 
\end{tabular}
\caption{The different benchmark points (BP's) for our study. These choices
  are made in view of the different scenarios that can arise in terms of
  production cross-sections, decay branching fractions and kinematic
  distributions. We also present the heavy quark pair production cross-section
  $\sigma_{q_H q_H}$ for the sum of all parton-level processes from
  Eqs.~(\ref{qHbarqH}) and (\ref{other_processes}) at the LHC with center of
  mass energies of $10 \tev$ and $14\tev$. }
\label{tab:BPs}
\end{center}
\end{table}

In Table~\ref{tab:BPs} we have also listed the total production cross-section
for the sum of all parton-level processes from the strong-interaction
processes from Eq.~(\ref{qHbarqH}) and the electroweak processes from
Eq.~(\ref{other_processes}) at the LHC with center of mass energies of $10
\tev$ and $14\tev$. For BP-2, the strong interaction processes dominate and
yield about $1.47\pb$, while the electroweak processes give only a
contribution of $0.60\pb$ at $14 \tev$, i.e.\ about 29\%. On the other hand,
for BP-3 with low $f=500\gev$ and therefore rather light $A_H, Z_H$ (exchanged
in the $t$-channel), the electroweak processes contribute $3.75\pb$ at a
center of mass energy of $14\tev$, i.e.\ 65\% of the total, compared to
$2.0\pb$ from strong interaction processes.  The biggest individual
parton-level cross-section is $\sigma(pp \to u_H d_H + X) = 2.39\pb$. This has
to be compared with the QCD process $\sigma(pp \to u_H \bar{u}_H + X) =
0.44\pb$ from $q\bar{q}$-annihilation and gluon-fusion, the latter
contributing about one third. The relative smallness of the QCD processes can
be understood from the small value of $\alpha_s(\mu_R = 2 m_{q_H} = 1.4\tev) =
0.084$ and the size of the various parton density functions for $\mu_F = 2
m_{q_H} = 1.4\tev$ around $x \sim m_{q_H} / 7\tev = 0.1$.


\section{Event generation: backgrounds and signal event selection}
\label{sec:event_generation}

As noted earlier, the initial parton level hard-scattering matrix elements in
the LHT with T-parity violation were calculated and the events generated with
the help of CalcHEP.  These events, along with the relevant masses, quantum
numbers and two-body and three-body decay branching fractions were passed on
to \PYTHIA~(Version 6.421)~\cite{Pythia} with the help of the SUSY Les Houches
Accord (SLHA) (v1.13) SUSY/BSM spectrum interface~\cite{SLHA_SUSY_BSM} for
their subsequent decays, showering, and hadronization. Initial and final state
radiations from QED and QCD and multiple interactions were also taken into
account in \PYTHIA. The SM backgrounds, except for $Z^{(*)}/\gamma^*$ (Drell-Yan process), were simulated with ALPGEN~\cite{ALPGEN} and then subsequently the unweighted event samples are passed onto \PYTHIA~ for their showering and hadronization. The matching of matrix-element hard partons and shower-generated jets is performed using the MLM prescription~\cite{ALPGEN}. This jet-parton matching allows us to generate inclusive samples of arbitrary jet multiplicity without any over-counting. Owing to the very large cross-section for the Drell-Yan process, it was not possible to generate statistically significant samples of events with additional hard jets using ALPGEN. Instead, the Drell-Yan background has been simulated with \PYTHIA. As we will explain below, after appropriate cuts, the Drell-Yan background is reduced to a negligible level. We have again used the leading order CTEQ6L parton distribution functions (for \PYTHIA~ we use the Les Houches Accord Parton Density Function (LHAPDF) interface~\cite{LHAPDF}). The QCD factorization and renormalization scales are in general kept fixed at the sum of the masses of the particles which are produced in the initial parton level scattering process. For the production of $Z^{(*)}/\gamma^*$, we have chosen the scale to be $M_Z$. If we would decrease the QCD scales by a factor of two, the cross-section can increase by about 30\%.

\subsection{Background from SM processes and within the LHT}
\label{sec:event_selection}

The main SM backgrounds for the dilepton channel come from the Drell-Yan
process via $Z^{(*)}/\gamma^*$ and the abundantly produced top quark pairs,
whereas for the four-lepton channel, $ZZ$ is the dominant source of
background.\footnote{Although $t\bar{t}$ events can also give rise to
four-lepton events, such backgrounds are relatively easily reduced with the requirement that among the four leptons, there are at least two opposite sign same flavor ones, whose invariant mass is around the $Z$ boson mass, followed  
by the effective mass cut, as shown in our subsequent analysis.} For
the dilepton channel we have also considered the backgrounds coming from $ZZ,
ZW$, $WW$ and $tW$. We have included additional multiple hard jets in the simulations as follows:  
\begin{itemize}
 \item $t\bar{t}+n$ jets, $0\leq n \leq 4 $
\item $ZZ+n$ jets, $0\leq n \leq 3 $
\item $ZW+n$ jets, $0\leq n \leq 3 $
\item $WW+n$ jets, $0\leq n \leq 3 $
\item $tW+n$ jets, $0\leq n \leq 1 $
\end{itemize}
 For the other possible dilepton and four-lepton backgrounds, we have checked that their cross-sections are small compared to the above ones, for instance, for $t \bar {t} Z$, where $Z \rightarrow l^{+}l^{-}$, the cross-section is around $40 \fb$ at $\sqrt{s}=14 \tev$.\footnote{In case of a jet faking  a lepton, $W+$~jets can also give rise 
to dilepton events. Although we do not consider the possibility of such fakes,
we expect the large $M_{eff}$ cut to reduce this background significantly.}
After putting a large effective mass cut as described below, these backgrounds are not expected to be significant. Additional leptons coming from the photons radiated by charged particles, or from the decay of pions, are generally expected to be removed by the basic isolation cuts described later.

For the strongly produced process $t\bar{t}~+$~jets, which has a long tail in the effective
mass distribution, we have multiplied the leading order cross-sections from
ALPGEN by appropriate K-factors wherever they are available in the literature. For $t\bar{t}~+$~0 jet the K-factor used is 2.2 from next-to-leading order (NLO) and next-to-leading-log resummed (NLL) corrections according to the analysis in
Ref.~\cite{Cacciari_et_al}. For $t\bar{t}~+$~1 jet we have used a K-factor of 1.29 according to the NLO calculation in Ref.~\cite{Dittmaier}, whereas for $t\bar{t}~+$~2 jets we used 1.28 as inferred from the recent NLO calculation in Ref.~\cite{Bevilacqua}.\footnote{Note that this K-factor has been obtained with a minimum $p_T$ of $50 \gev$ for the jets whereas we will employ a cut of only $20 \gev$. Based on the observation in Ref.~\cite{Huston}, that in many cases the K-factor diminishes for processes with more hard jets, the actual K-factor for $t \bar{t}~+$~2 jets might be lower than 1.28.}

In addition, two-and four-lepton final states can occur from processes within
the LHT model itself. The heavy T-odd gauge bosons, $W_H$ and $Z_H$, produced
in various cascades, can lead to hard isolated leptons. Leptons from b-quarks
and $\tau$'s in LHT cascades can also fake our signals {\it prima facie}. For
small $\kappa_l$, T-odd leptons $l_{H}$ will also give rise to leptons in the
final state. Since we have taken $\kappa_l = 1$, the latter decay does not
occur for the chosen benchmark points. On the whole, substantial as several of
the aforementioned backgrounds may be, they do not in general affect 
the invariant mass peaks from $A_H$ decays.

\subsection{Event selection criteria}

In our analysis we demand for the {\it dilepton signal} that we have exactly
one pair of opposite sign same flavor (OSSF) leptons from the decay $A_H \to
l^+ l^-$, with $l=\{e,\mu\}$. For the {\it four-lepton signal} from the decay
$A_H \to ZZ^{(*)} \to l^+ l^- {l^\prime}^+ {l^\prime}^-$, where $l,
l^\prime=\{e,\mu\}$, we demand that there should be four leptons, among which
at least one OSSF lepton pair should have an invariant mass peaked around
$M_Z$ (i.e.,\ $M_Z-20 \gev \leq M_{ll} \leq M_Z+20 \gev$). The last criterion
is used because in the scenarios that we consider, at least one $Z$-boson is
on-shell.

The following basic selection cuts (denoted by Cut-1 below) were applied for
both the signal and the background~\cite{TDR,Mellado}: 

\noindent
{\bf Lepton selection:}
\begin{itemize}
 \item $p_T>$ 10 GeV and $|\eta_{\ell}|<$ 2.5, where $p_T$ is
   the transverse momentum and $\eta_{\ell}$ is the pseudorapidity of the
   lepton (electron or muon). 
\item {\bf Lepton-lepton separation:}  ${\Delta R}_{\ell\ell} \geq $ 0.2,
  where $\Delta R = \sqrt {(\Delta \eta)^2 + (\Delta \phi)^2}$ is the
  separation in the pseudorapidity--azimuthal angle plane.
\item {\bf Lepton-jet separation:} $\Delta R_{\ell j} \geq 0.4$ for all jets
  with $E_T >$ 20 GeV.  
\item The total energy deposit from all {\it hadronic activity} within a cone
  of $\Delta R \leq 0.2$ around the lepton axis should be $\leq$ 10 GeV. 
\end{itemize}

\noindent
{\bf Jet selection:}

\begin{itemize}
\item 
Jets are formed with the help of {\tt PYCELL}, the inbuilt cluster routine in
\PYTHIA.  The minimum $E_{T}$ of a jet is taken to be $20\gev$, and
we also require  $|\eta_j|<$ 2.5. 
\end{itemize}

We have approximated the detector resolution effects by smearing the energies
(transverse momenta) with Gaussian functions. The different contributions to
the resolution error have been added in quadrature.

\begin{itemize}
\item {\bf Electron energy resolution:}
\be
\frac{\sigma(E)}{E} = \frac{a}{\sqrt{E}} \oplus b \oplus
                      \frac{c}{E}, 
\ee
where
\be
(a, b, c) = \left\{ \begin{array}{lcl}
                   (0.030\gev^{1/2}, \, 0.005, \, 0.2\gev), & \hspace{1em} &
                                                   |\eta| < 1.5, \\ 
                    (0.055\gev^{1/2}, \, 0.005, \, 0.6\gev), & \hspace{1em} &
						   1.5 < |\eta| < 2.5. 
		    \end{array}
            \right.
\ee
\item {\bf Muon $p_T$ resolution:}
\be
\frac{\sigma(P_T)}{P_T} = \left\{ \begin{array}{lcl}
                       a , & \hspace{1em} &  p_{T} < 100\gev, \\
                       \displaystyle
                         a + b \, \log \frac{p_T}{100\gev} , & & 
			 p_{T}>100\gev, 
		    \end{array}
            \right.
\ee
with
\be
(a, b) = \left\{ \begin{array}{lcl}
                       (0.008, \, 0.037), & \hspace{1em} & |\eta| < 1.5,  \\
                       (0.020, \, 0.050), & \hspace{1em} & 1.5 < |\eta| <
                                                           2.5. \\  
		    \end{array}
            \right.
\ee
\item {\bf Jet energy resolution:}
\be
\frac{\sigma(E_T)}{E_T} = \frac{a}{\sqrt{E_T}}, 
\ee
with 
$ a= 0.5\gev^{1/2}$, the default value used in {\tt PYCELL}.
\end{itemize}

Under realistic conditions, one would of course have to deal with
aspects of misidentification of leptons.

An important cut will be imposed on the effective mass variable, defined to be
the scalar sum of the transverse momenta of the isolated leptons and jets and
the missing transverse energy, 
\be \label{M_eff}
 M_{eff}=\sum p_{T}^{jets} +\sum p_{T}^{leptons} + \met,  
\ee 
where the missing transverse energy is given by 
\be \label{missing_E_T}
\met =\sqrt{\left(\sum p_x \right)^2+\left(\sum p_y\right)^2}. 
\ee
Here the sum goes over all the isolated leptons, the jets, as well as the
`unclustered' energy deposits. The energies of the `unclustered' components,
however, have not been smeared in this analysis.

\begin{figure}[h!]
\begin{center}
\centerline{\epsfig{file=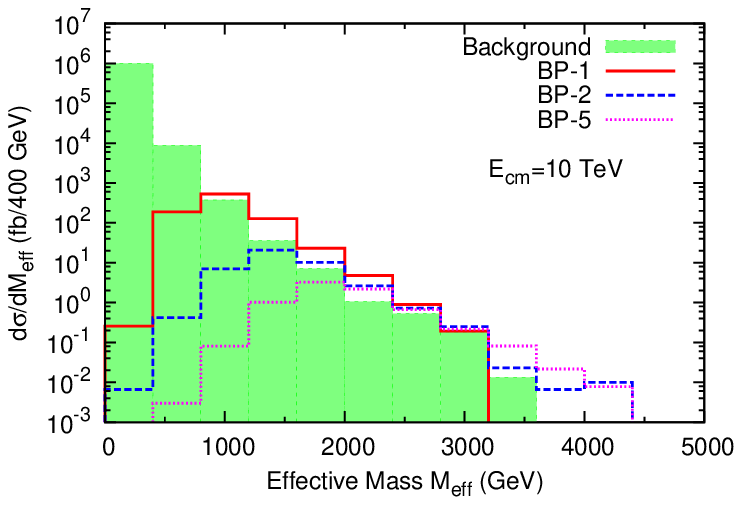,width=12.0cm,height=9.0cm,angle=-0}} 

\vspace*{0.25cm}
\centerline{\epsfig{file=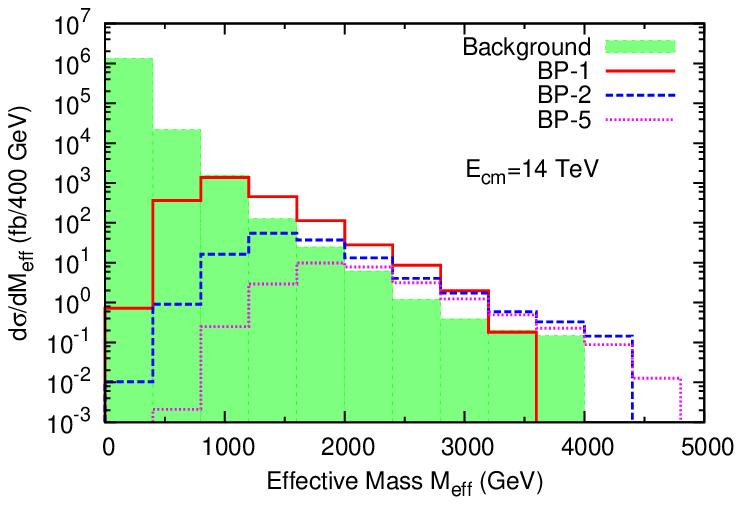,width=12.0cm,height=9.0cm,angle=-0}} 

\caption{Effective mass distribution of dilepton events after the basic cuts
   (Cut-1) for BP-1 ($m_{q_H} \sim 400\gev$), BP-2 ($m_{q_H} \sim 700\gev$),
   BP-5 ($m_{q_H} \sim 1000\gev$) and the SM background, mostly Drell-Yan and
   $t\bar{t}~+$~jets, at $\sqrt{s}=10\tev$ (top panel) and $\sqrt{s}=14\tev$ (bottom 
   panel).}
\label{fig:Meff}
\end{center}
\end{figure}

In Fig.~\ref{fig:Meff} we plot the distribution of the effective mass after
  the basic cuts (Cut-1) for dilepton events for the benchmark points BP-1
  ($m_{q_H} \sim 400\gev$), BP-2 ($m_{q_H} \sim 700\gev$), BP-5 ($m_{q_H} \sim
  1000\gev$) and the SM background, dominantly from Drell-Yan and
  $t\bar{t}~+$~jets. Recall that we have included appropriate K-factors for the latter process. The SM backgrounds are huge; however, the
  distributions peak around $2 m_{q_H}$ for the signal and between $M_Z$ and
  $2 m_t$ for the SM background. It is clear that the production of heavy
  particles in the initial hard-scattering will lead to a high $M_{eff}$ in
  most cases.  Therefore, imposing the {\it effective mass cut:}
\be \label{M_eff_cut}
M_{eff} \geq 1 \tev \qquad (\mbox{Cut-2}), 
\ee
will reduce the SM background substantially, although the distribution from SM processes with additional hard jets has a long tail towards larger values of $M_{eff}$. Note that the
log-plot makes the differences between the curves look smaller than they
actually are. Although this fixed cut also reduces the signal for the lighter
$m_{q_H} \sim 400\gev$ from BP-1 by about half, the corresponding larger
production cross-section makes up for this loss. In a more realistic analysis
one would of course try to optimize the choice of the cut on $M_{eff}$,
depending on the expected signal.

In addition to the effective mass cut from Eq.~(\ref{M_eff_cut}), we will also
use the fact that we expect a peak in the invariant mass distribution of
dileptons ($M_{ll}$) or four leptons ($M_{4l}$) near the mass of $A_H$. In our
analysis, we will therefore impose the condition that the invariant mass
should be in a {\it window around $M_{A_H}$}
\be \label{M_ll_cut}
M_{A_{H}}-20 \gev \leq M_{ll, 4l} \leq M_{A_{H}}+20 \gev \qquad
(\mbox{Cut-3}). 
\ee
As we will see, this condition will in particular help to reduce the
background from leptons within the LHT model. Of course, in a realistic
experimental analysis, where the mass of $A_H$ is unknown, one would scan the
whole range of invariant masses and impose such a window around some seed-mass
$M_{A_H}$, thereby looking for an excess of the signal over the SM and LHT
model backgrounds, which are almost flat except near the $Z$-mass.  Moreover,
this excess should stand out in this window, as compared to the adjoining
bins.

Unfortunately, the need to implement such a cut in our analysis will not allow
us to detect a heavy photon $A_H$ with a mass very close to $M_Z$, since in
that case the cut cannot free the peak from contamination by $Z$-production
within the SM and in LHT cascades. In principle, by looking at the relative size of the
branching fractions into charged leptons, including $\tau$, and hadrons
(jets), one might still be able to distinguish the $A_H$ from the
$Z$-boson. However, if we look at a heavy photon with $M_{A_H} = 92\gev$, we
get $R_{A_H} \equiv \BR(A_H \to \mbox{quarks}) / \BR(A_H \to \mbox{all charged
leptons}) = 6.27$, which is not much different from the corresponding ratio
for the $Z$-boson, $R_Z = 6.92$. So the signature of $A_H$ in this situation
might be difficult to distinguish at the LHC. On the other hand, if the
effective couplings of all charged leptons with the $A_H$ are scaled up by a
factor of two, as discussed above, we would get $R_{A_H} = 1.57$, which is
quite different from the value for the $Z$-boson. Note that in this case
$\BR(A_H \to e^+ e^-) + \BR(A_H \to \mu^+ \mu^-)$ changes from $7.58\%$ for
the original couplings to $22.66\%$ for the rescaled leptonic couplings, i.e.\
it increases by a factor three and the required luminosity decreases
correspondingly.


\section{Results}
\label{sec:results}

\subsection{Dilepton signal}

\subsubsection{LHC with $\sqrt{s} = 10\tev$}
\label{subsubsec:Dilepton_10-TeV}

In Table~\ref{tab:Background-10TeV} we list, for the opposite sign
same flavor (OSSF) dilepton signal ($l=e,\mu$), the cross-sections of
the dominant SM background processes after the basic cuts (Cut-1) and the
effective mass cut $M_{eff} > 1\tev$ (Cut-2) for the LHC running at
$\sqrt{s} = 10\tev$.  We can see that the effective mass cut reduces
all the SM dilepton backgrounds significantly, in particular from the
Drell-Yan process via $Z/\gamma^*$, which is the overwhelming
background after the basic cuts. After the cut on $M_{eff}$,
$t\bar{t}~+$~jets is the largest background due to the long tail in the
effective mass distribution, see Fig.~\ref{fig:Meff}.\footnote{We should note that in the simulation with ALPGEN a substantial number of events in $t \bar{t}~+$~4 jets and $VV~+$~3 jets ($V=W,Z$) pass the $M_{eff}$ cut. We can therefore not exclude the possibility that the inclusion of more hard jets might increase the total background cross-section after Cut-2 by some amount. Such a simulation is beyond the scope of our present study. At least part of the effects of these hard jets is taken into account by the PYTHIA showering and MLM matching of the ALPGEN samples with the highest jet multiplicity.}
\begin{table}[htb]
\begin{center}
\begin{tabular}{|c|r|r|}
\hline
Background & \multicolumn{1}{c|}{Cut-1} & \multicolumn{1}{c|}{Cut-2} \\ 
& \multicolumn{1}{c|}{(fb)} &\multicolumn{1}{c|}{(fb)}\\
\hline
$Z/\gamma^{*}$ &1247174 &$\sim$ 0.00 \\
\hline 
$t\bar{t}~+$~jets &6278 &84.03 \\
\hline
$ZZ~+$~jets &546 &5.76 \\
\hline
$WW~+$~jets &946 &9.58  \\
\hline
$ZW~+$~jets &624 &13.39 \\ 
\hline
$tW~+$~jets &719 &8.37  \\
\hline \hline 
Total& 1256287& 121.13 \\
\hline 
\end{tabular}
\caption{Dominant opposite sign same flavor dilepton ($l= e,\mu$) SM
    background cross-sections for $\sqrt{s}=10\tev$ after the basic cuts
    (Cut-1) and after the cut $M_{eff} \geq 1\tev$ (Cut-2).}
\label{tab:Background-10TeV}
\end{center}
\end{table}

In our simulation of the Drell-Yan process with \PYTHIA, out of $10^6$ Monte-Carlo (MC) events, we did not see any dilepton event with $M_{eff} > 1\tev$ . Actually,
in the simulation for the LHC running at $14\tev$, where the cross-section is
higher and where we expect more events with larger $M_{eff}$, we did not get a
single event with $M_{eff} > 1\tev$ out of $10^7$ simulated Drell-Yan
events. A larger MC sample would be needed to put a definite number on the
cross-section after Cut-2. If we simply assume an upper bound of one event after Cut-2,
which is probably much bigger than the correct number, this leads to an upper
bound on the dilepton cross-section from $Z/\gamma^*$ after Cut-2 of about
$45\fb$, i.e.\ quite sizeable compared to $t\bar{t}$.  In the following we
assume that we can neglect the SM background from $Z/\gamma^*$ after the
effective mass cut. In any case, after the Cut-3, i.e.\ that the dilepton
invariant mass should be in a narrow window around $M_{A_H}$, see
Eq.~(\ref{M_ll_cut}), a further reduction of the cross-section and number of
dilepton events from $Z/\gamma^*$ will occur anyway.

\begin{figure}[h!]
\begin{center}
\vspace*{-0.5cm}
\centerline{\epsfig{file=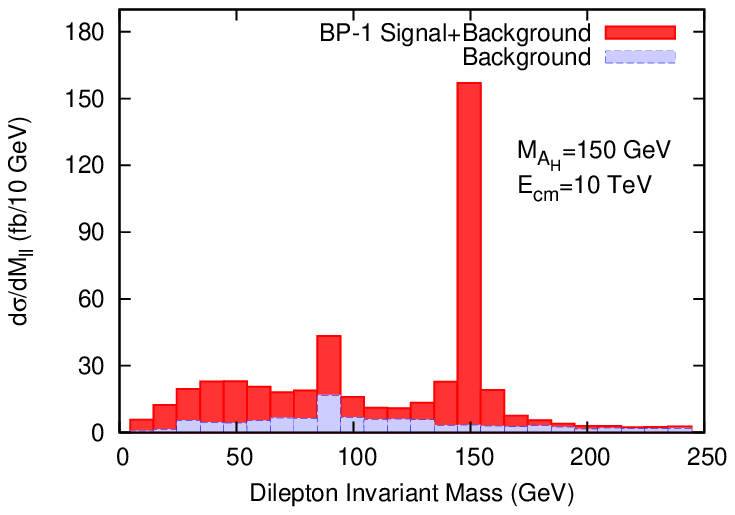,width=7.5cm,height=5.7cm,angle=-0}
\hskip 20pt \epsfig{file=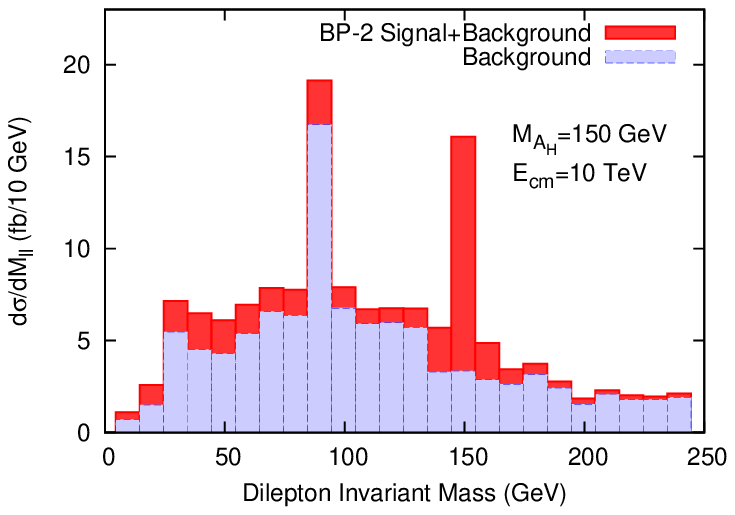,width=7.5cm,height=5.7cm,angle=-0}}
\centerline{\epsfig{file=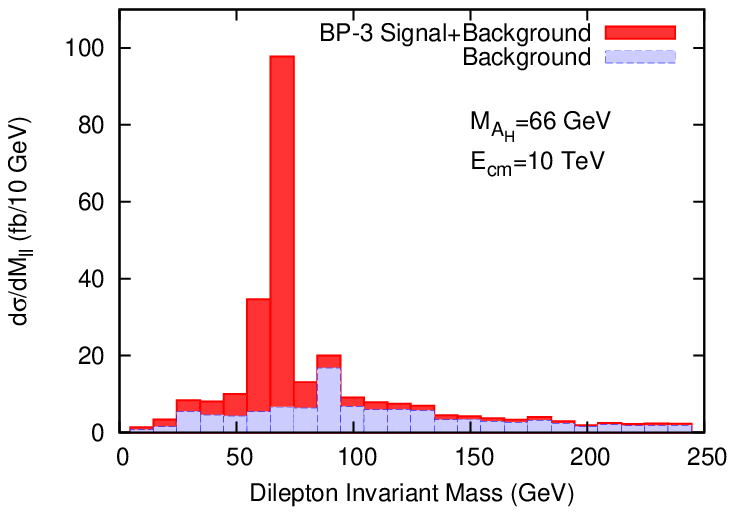,width=7.5cm,height=5.7cm,angle=-0}
\hskip 20pt \epsfig{file=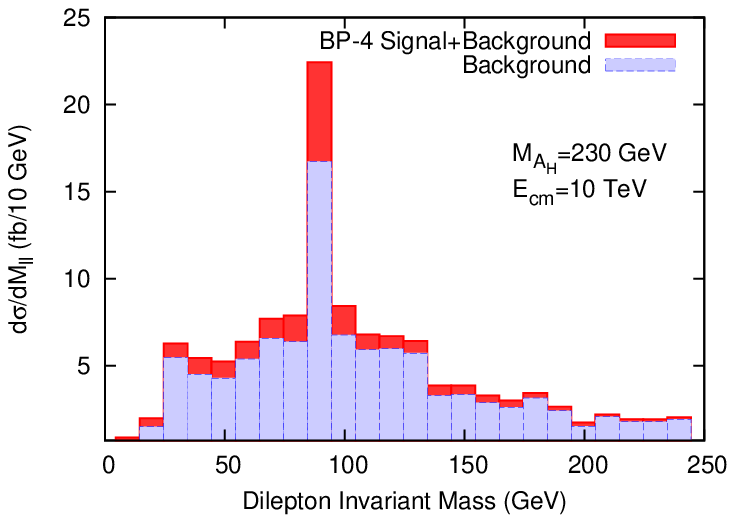,width=7.5cm,height=5.7cm,angle=-0}}
\centerline{\epsfig{file=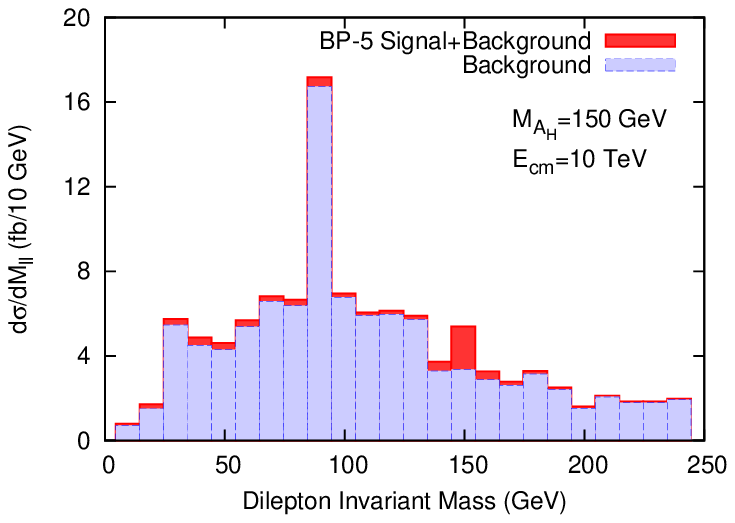,width=7.5cm,height=5.7cm,angle=-0}
\hskip 20pt \epsfig{file=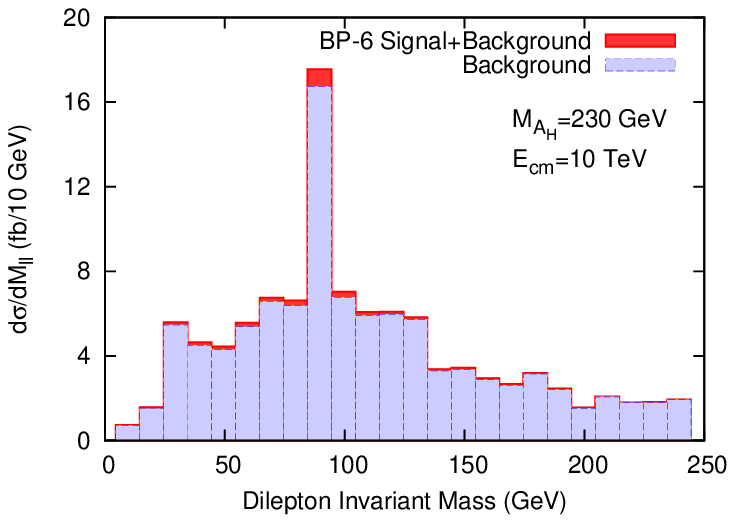,width=7.5cm,height=5.7cm,angle=-0}}
\caption{Invariant mass distribution of OSSF dilepton pairs ($l=e,\mu$) for
  $\sqrt{s}=10\tev$ after the basic cuts (Cut-1) and the effective mass cut
  $M_{eff} \geq 1\tev$ (Cut-2) for all the benchmark points, with the
  corresponding SM background. Note that the `signal' also includes
  dileptons within the LHT not coming from $A_H$, but, for instance through
  the decay of the $Z$-boson. This is in particular the case for BP-4 and BP-6
  where we do not expect any dileptons from $A_H$. For an integrated
  luminosity of $200\pb^{-1}$ there are 73 signal events for BP-1 and 33
  events for BP-3, but only 8 signal events for BP-2 and 1.5 events for BP-5,
  compared to 24 events from the SM background. See Table~\ref{tab:OSSF-10TeV}
  for more details.}
\label{fig:inv_mass_dilepton_10TeV}
\end{center}
\end{figure}

In Fig.~\ref{fig:inv_mass_dilepton_10TeV} we plot the invariant mass
distributions of OSSF dileptons for all the benchmark points and the
corresponding SM background after the basic cuts (Cut-1) and the $M_{eff}$ cut
(Cut-2). For BP-1, BP-2, BP-3 and BP-5, a peak emerges at the mass of $A_H$,
in particular very clearly for BP-1 and BP-3. BP-1 has a large parton-level
T-odd quark pair production cross-section because of the relatively small
$m_{q_H} \sim 400\gev$, compared to $m_{q_H} \sim 700\gev$ for BP-2. BP-3 has
also $m_{q_H} \sim 700\gev$, but a small value of $f=500\gev$. Therefore the
T-odd quark production cross-section is strongly enhanced by electroweak
contributions, as discussed earlier, see Table~\ref{tab:BPs}. Furthermore, the
leptonic branching ratio for the light $M_{A_H} = 66\gev$ is also large, see
Table~\ref{tab:A_H-Decay}, leading to more dilepton events. Note, however,
that the `signal' after Cut-2 also includes dileptons from within the LHT
coming for instance from the decay via the $Z$-boson, leading to an enhancement in the peak at $M_Z$ for BP-1 and BP-2 in
Fig.~\ref{fig:inv_mass_dilepton_10TeV}. Since for BP-3 with $f=500\gev$ the
branching ratio $A_H \to l^+ l^-$ is higher than for BP-1 and BP-2, the
$Z$-peak is much smaller compared to the $A_H$-peak.

The total integrated luminosity at the LHC running at $10\tev$ will probably
be around $200\pb^{-1}$.  For this integrated luminosity, there will be 73
dilepton `signal' events for BP-1 and 33 events for BP-3, compared to a SM
background of 24 events. Note, however, that there will be only 8 dilepton
events for BP-2 and 1.5 events for BP-5.

As mentioned in Section~\ref{sec:T_violation}, the total decay width of $A_H$
is only of the order of eV and therefore radiative corrections and detector
effects will determine the observed width of the resonance peak in reality.

For BP-4 and BP-6 with $f=1500\gev$, the decay branching ratio $A_H \to l^+
l^-$ is negligible compared to $A_H \to Z Z^{(*)}$, see
Table~\ref{tab:A_H-Decay}, and therefore the four-lepton signal will be the
relevant signature for discovery. Nevertheless, we get some dilepton events,
in particular for BP-4, but the dilepton invariant mass distribution peaks at
the $Z$-boson mass and not at $M_{A_H}$.

In order to reduce the SM background further, and also to eliminate the
background from dileptons within the LHT which do not come from the decay $A_H
\to l^+ l^-$, we impose the additional constraint that the invariant mass of
the dileptons should be in a window of $\pm 20\gev$ around $M_{A_H}$ (Cut-3),
see Eq.~(\ref{M_ll_cut}).  In Table~\ref{tab:OSSF-10TeV} we give the
cross-sections for OSSF dilepton events after the basic cuts (Cut-1) and after
Cut-2 and Cut-3 for all the benchmark points for the LHC running at $\sqrt{s}
= 10\tev$. The total cross-section for the SM background after Cut-2 and Cut-3
is also given. We also list in Table~\ref{tab:OSSF-10TeV} the number of signal
and background events after Cut-2 and after Cut-3 for an integrated luminosity
of $200\pb^{-1}$.

\begin{table}[tbh]
\begin{center}
\begin{tabular}{|c|r|r|c|r|c|r|r|r|r|}
\hline
&\multicolumn{1}{c|}{Cut-1} & \multicolumn{1}{c|}{Cut-2} &
\multicolumn{1}{c|}{Cut-2}  & \multicolumn{1}{c|}{$S_2$} &
\multicolumn{1}{c|}{$B_2$} & \multicolumn{1}{c|}{Cut-3} &
\multicolumn{1}{c|}{Cut-3} &\multicolumn{1}{c|}{$S_3$} &
\multicolumn{1}{c|}{$B_3$}  \\  
&\multicolumn{1}{c|}{[S]} & \multicolumn{1}{c|}{[S]} &
\multicolumn{1}{c|}{[BG]}& & & \multicolumn{1}{c|}{[S]} &
\multicolumn{1}{c|}{[BG]}& & \\ 
&\multicolumn{1}{c|}{(fb)} &\multicolumn{1}{c|}{(fb)} &
     \multicolumn{1}{c|}{(fb)} &  &  & \multicolumn{1}{c|}{(fb)} &
     \multicolumn{1}{c|}{(fb)} &  &  \\  
\hline
BP-1 & 871.5 & 367.0 & 121.1 & 73.4 & 24.2 & 196.7 & 12.4 & 39.3 &2.5 \\
\hline
BP-2 & 41.9  & 39.8  & 121.1 & 8.0  & 24.2 & 18.2  & 12.4 & 3.6  &2.5 \\
\hline
BP-3 & 175.2 & 168.1 & 121.1 & 33.6 & 24.2 & 132.9 & 23.3 & 26.6 & 4.7 \\
\hline
BP-4 & 21.9  & 21.3  & 121.1 & 4.3  & 24.2 & 0.5   & 7.2  & 0.1  &1.4\\
\hline 
BP-5 & 7.6   & 7.6   & 121.1 & 1.5  & 24.2 & 3.1   & 12.4 & 0.6  & 2.5\\
\hline 
BP-6 & 3.5   & 3.5   & 121.1 & 0.1  & 24.2 & 0.1   & 7.2  & 0.02 &1.4 \\ 
\hline 
\end{tabular}
\caption{OSSF dilepton signal (S) from all T-odd quark-pair production
    processes and total SM background (BG) cross-sections at the LHC with
    $\sqrt{s}=10\tev$ after the different cuts described in the text. The
    number of signal and background events after Cut-2 with an integrated
    luminosity of $200\pb^{-1}$ are also given as $S_2$ and $B_2$,
    respectively. The corresponding numbers after Cut-3 are denoted by $S_3$
    and $B_3$.}
\label{tab:OSSF-10TeV}
\end{center}
\end{table}

We can see from Table~\ref{tab:OSSF-10TeV} that with an integrated luminosity
of $200\pb^{-1}$, only BP-1 and BP-3 yield a clear signal over the SM
background after Cut-2. The Cut-3 then reduces the SM background and the
dilepton background from within the LHT model almost completely, with $S_3/B_3
= 15.7$ for BP-1 and $S_3/B_3 = 5.7$ for BP-3 with more that 10 signal events
for both benchmark points. Although the high $M_{eff} \geq 1\tev$ cut
(Cut-2) reduces the signal for BP-1 with a low mass $m_{q_H} = 400\gev$ by
about a factor two, the larger production cross-section compensates for that.
Therefore, at least for BP-1 and BP-3, one expects a clear dilepton signal 
from the decay of the $A_H$ in the LHT
with T-parity violation at the early stage of the LHC run with low center of
mass energy and modest luminosity. With the predicted number of signal events
after Cut-3, it will presumably also be possible to reconstruct the mass of
$A_H$ and determine the symmetry breaking scale $f$. 

If we demand that we have at least 10 signal events, BP-2 yields not enough
events with an integrated luminosity of $200\pb^{-1}$, in particular after
Cut-3. Furthermore, with this luminosity there will be almost no dilepton
events for BP-5, already after Cut-2, because of the small production
cross-section for $m_{q_H} \sim 1\tev$. Also the branching ratio of $A_H$ into
dileptons is small for this benchmark point, since $f=1\tev$.

Note that the BP-4 with $M_{A_H} = 230\gev$ yields about half the dilepton
events of BP-2 for the same mass $m_{q_H} \sim 700\gev$ of the heavy T-odd
quarks.  However, these dileptons for BP-4 are not coming from the decay $A_H
\to l^+ l^-$, but from other sources, mostly the $Z$-boson, as mentioned
above. In Table~\ref{tab:OSSF-10TeV} this is visible after imposing the Cut-3
which almost completely removes all dilepton events for BP-4, whereas about
half the dilepton events survive for BP-2. For BP-6 there are essentially no
dilepton events for $200\pb^{-1}$. As for BP-4, we expect for this benchmark
point, which has $f=1500\gev$ and $M_{A_H} = 230\gev$, the four-lepton mode to
be relevant for discovery.

We should caution the reader about the numbers given in
Table~\ref{tab:OSSF-10TeV} for the SM background cross-sections and the number
of BG events after Cut-2 and in particular after Cut-3. The total number of MC events in the OSSF channel we simulated (including all possible processes) is 
350987 after Cut-1. After the cut on $M_{eff}$ (Cut-2), we have 4296
events in our MC sample and after Cut-3 there remain 644 MC events in the
window of $\pm 20\gev$ around the $A_H$ mass (for the case $M_{A_H} =
66\gev$). Therefore, there is an intrinsic uncertainty of about 4\% on the
numbers for $B_3$ given in the Table~\ref{tab:OSSF-10TeV}. A much larger MC
simulation to pin down these numbers more precisely is beyond the scope of the
present work. Note, however, that we have enough simulated events for the
signal. For instance, for BP-2, we have 12041 MC events after Cut-2 and 5501
MC events after Cut-3.

\subsubsection{LHC with $\sqrt{s} = 14\tev$}

In Table~\ref{tab:Background-14TeV} we list, for the opposite sign same flavor
(OSSF) dilepton signal ($l=e,\mu$), the cross-sections of the dominant SM
background processes after the basic cuts (Cut-1) and the effective mass cut
$M_{eff} > 1\tev$ (Cut-2) for the LHC running at $\sqrt{s} = 14\tev$. As for
$10\tev$, we can see that the effective mass cut reduces all the SM dilepton
backgrounds significantly, in particular from the Drell-Yan process via
$Z/\gamma^*$ which is the main background after the basic cuts. Again, after
the cut on $M_{eff}$, $t\bar{t}$ is the largest background due to the long
tail in the effective mass distribution, see Fig.~\ref{fig:Meff}.  As noted in
the previous subsection, in our simulation of the Drell-Yan process with \PYTHIA, we did not see any
dilepton event with $M_{eff} > 1\tev$ out of $10^7$ MC events. There
were 30048 OSSF dilepton MC events which passed the basics cuts. If we would
simply assume an upper bound of one event after the Cut-2, which might be way
off the correct number, this would lead to an upper bound on the dilepton
cross-section from $Z/\gamma^*$ after Cut-2 of about $57\fb$, i.e.\ again
quite sizeable compared to $t\bar{t}$.  In the following we assume again that
we can neglect the SM background from $Z/\gamma^*$ after Cut-2.

\begin{table}[htb]
\begin{center}
\begin{tabular}{|c|r|r|}
\hline
Background & \multicolumn{1}{c|}{Cut-1} & \multicolumn{1}{c|}{Cut-2} \\ 
& \multicolumn{1}{c|}{(fb)} &\multicolumn{1}{c|}{(fb)}\\
\hline
$Z/\gamma^*$ &1711746 &$\sim$ 0.00 \\
\hline
$t\bar{t}~+$~jets &13854 &344.33  \\
\hline
$ZZ~+$~jets &827 &13.82 \\
\hline
$WW~+$~jets &1385 &30.17 \\
\hline
$ZW~+$~jets &951 &43.40  \\
\hline
$tW~+$~jets &1604&41.89  \\
\hline\hline 
Total&1730366 & 473.61 \\
\hline
\end{tabular}
\caption{Same as Table~\ref{tab:Background-10TeV} for $\sqrt{s}=14\tev$.}
\label{tab:Background-14TeV}
\end{center}
\end{table}

\begin{figure}[h!]
\begin{center}
\centerline{\epsfig{file=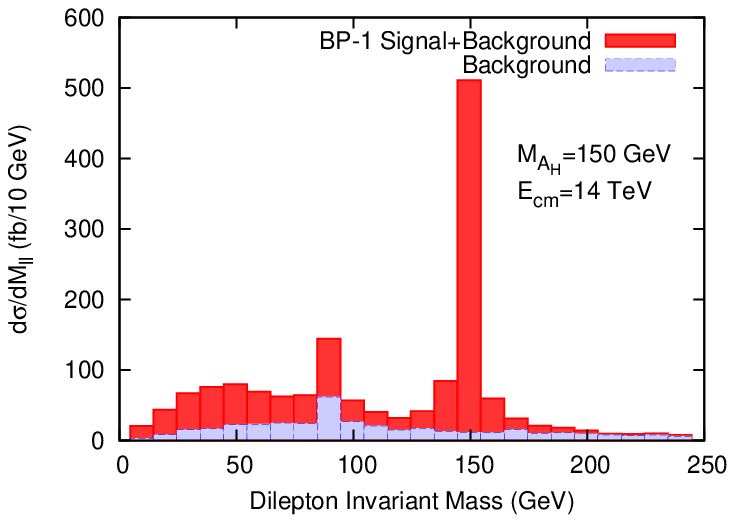,width=7.5cm,height=5.85cm,angle=-0}
\hskip 20pt \epsfig{file=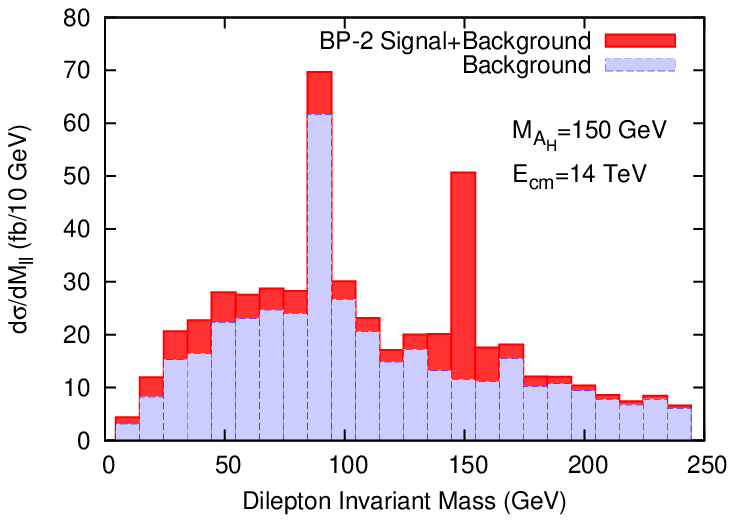,width=7.5cm,height=5.85cm,angle=-0}}
\centerline{\epsfig{file=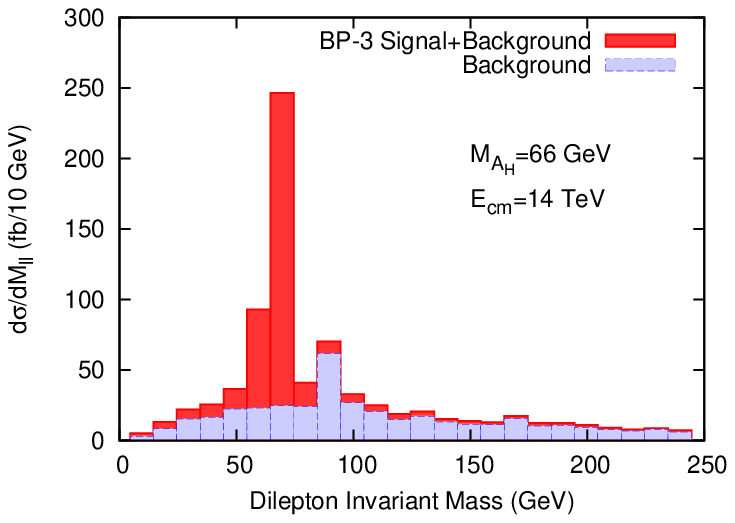,width=7.5cm,height=5.85cm,angle=-0}
\hskip 20pt \epsfig{file=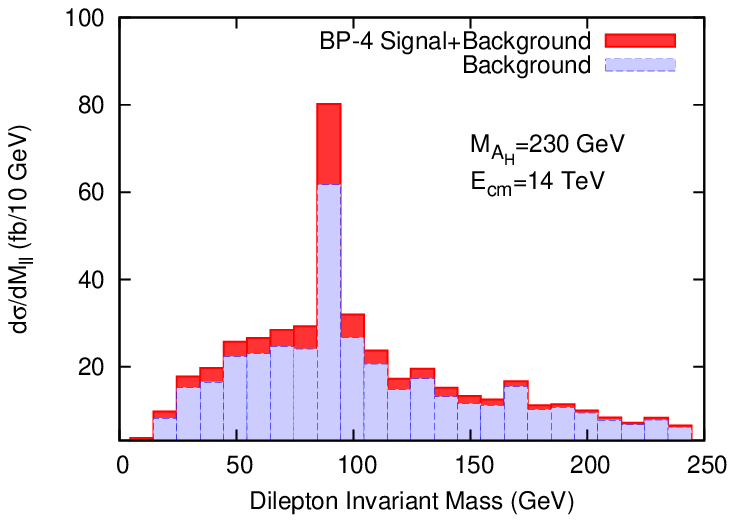,width=7.5cm,height=5.85cm,angle=-0}}
\centerline{\epsfig{file=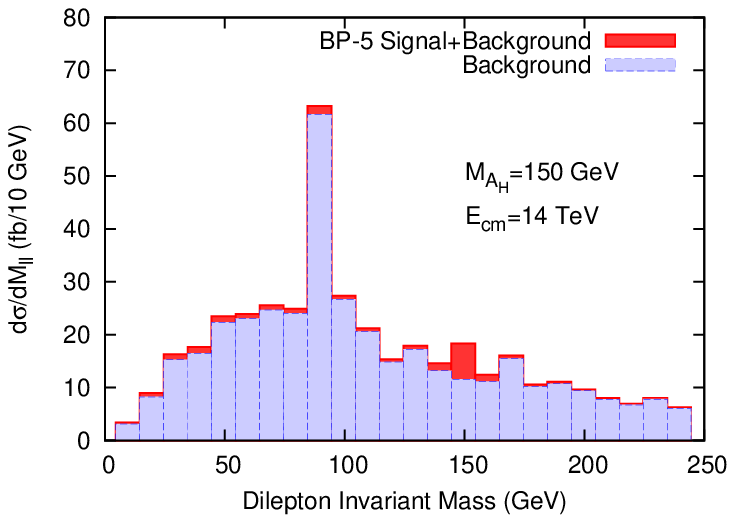,width=7.5cm,height=5.85cm,angle=-0}
\hskip 20pt \epsfig{file=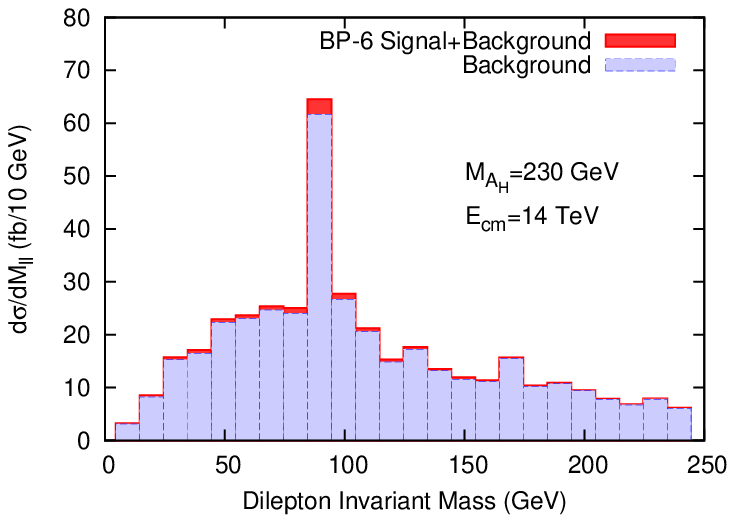,width=7.5cm,height=5.85cm,angle=-0}}
\caption{Same as Fig.~\ref{fig:inv_mass_dilepton_10TeV} for
  $\sqrt{s}=14\tev$.} 
\label{fig:inv_mass_dilepton_14TeV}
\end{center}
\end{figure}

In Fig.~\ref{fig:inv_mass_dilepton_14TeV} we plot the invariant mass
distributions of OSSF dileptons for all the benchmark points and the
corresponding SM background after the basic cuts (Cut-1) and the cut on
$M_{eff}$ (Cut-2). Again for BP-1, BP-2, BP-3 and BP-5, a peak emerges at the
mass of $A_H$, in particular very clearly for BP-1, BP-2 and BP-3.  With an
integrated luminosity of $30\fb^{-1}$, there are 36088 signal events for BP-1,
3723 events for BP-2 and 2135 events for BP-3, compared to 14208 SM background
events.  Again, for BP-1 and BP-2 we have a non-negligible amount of dilepton
background from within the LHT, in particular from the decays via the
$Z$-boson, as can be clearly seen in the
Fig.~\ref{fig:inv_mass_dilepton_14TeV}.

As mentioned before, for BP-4 and BP-6 with $f=1500\gev$, the decay branching
ratio $A_H \to l^+ l^-$ is negligible compared to $A_H \to Z Z^{(*)}$, see
Table~\ref{tab:A_H-Decay}, and therefore the four-lepton signal will be the
relevant signature for discovery. Nevertheless, we get many dilepton events
even for these two benchmark points, but there is a peak at the $Z$-boson mass
and not at $M_{A_H}$.

In order to reduce the SM background further, but also to eliminate the
background from dileptons within the LHT which do not come from the decay $A_H
\to l^+ l^-$, we impose again the additional constraint that the invariant
mass of the dileptons should be in a window around $M_{A_H}$ (Cut-3), see
Eq.~(\ref{M_ll_cut}).  In Table~\ref{tab:OSSF-14TeV} we give the
cross-sections for OSSF dilepton events after the basic cuts (Cut-1) and after
Cut-2 and Cut-3 for all the benchmark points for the LHC running at $\sqrt{s}
= 14\tev$. The total cross-section for the SM background after Cut-2 and Cut-3
is also given. We also list in Table~\ref{tab:OSSF-14TeV} the number of signal
and background events after Cut-2 and after Cut-3 for an integrated luminosity
of $30\fb^{-1}$.

The qualitative features of the benchmark points for the LHC running at
$\sqrt{s} = 14\tev$ are very similar to the case of $\sqrt{s} = 10\tev$, but
now we have higher event rates. First of all, after the Cut-2, all the signal
and background cross-sections are about a factor of three bigger, see
Tables~\ref{tab:OSSF-10TeV} and \ref{tab:OSSF-14TeV}. Furthermore, we assume
that we have now much more integrated luminosity, $30\fb^{-1}$ compared to
$200\pb^{-1}$ earlier. The Cut-2 again removes about half of the signal events
for BP-1.  For all benchmark points, we now get more than 10 signal events
even after Cut-3 and also the number of background events is much larger than
100. It makes therefore sense to consider the signal significance by looking
at $S/\sqrt{B}$ which is also given in Table~\ref{tab:OSSF-14TeV} for the
number of events after Cut-3.

\begin{table}[h!]
\begin{center}
\begin{tabular}{|c|r|r|c|r|c|r|r|r|r|r|}
\hline
&\multicolumn{1}{c|}{Cut-1} & \multicolumn{1}{c|}{Cut-2} &
\multicolumn{1}{c|}{Cut-2} & \multicolumn{1}{c|}{$S_2$} &
\multicolumn{1}{c|}{$B_2$} & \multicolumn{1}{c|}{Cut-3} &
\multicolumn{1}{c|}{Cut-3} & \multicolumn{1}{c|}{$S_3$} &
\multicolumn{1}{c|}{$B_3$} &\multicolumn{1}{c|}{$S_3/\sqrt{\!B_3}$} \\ 
&\multicolumn{1}{c|}{[S]} & \multicolumn{1}{c|}{[S]} &
\multicolumn{1}{c|}{[BG]}& & & \multicolumn{1}{c|}{[S]} &
\multicolumn{1}{c|}{[BG]}& & &\\ 
&\multicolumn{1}{c|}{(fb)} &\multicolumn{1}{c|}{(fb)} &
     \multicolumn{1}{c|}{(fb)} & & & \multicolumn{1}{c|}{(fb)}
     &\multicolumn{1}{c|}{(fb)} & & & \\   
\hline
BP-1 & 2341.8 &1202.9 & 473.6&36088 &14208&642.9&50.3&19286&1508 & 496.6\\
\hline
BP-2 & 129.1 & 124.1 & 473.6&3723 &14208&55.5&50.3&1665&1508 & 42.9 \\
\hline
BP-3 & 428.9 & 413.3 &473.6& 12398 &14208&322.9&95.8&9686&2873 & 180.7 \\
\hline
BP-4 & 72.7 & 71.2 & 473.6&2135 &14208&1.9&27.0&56&809 & 2.0  \\
\hline 
BP-5 & 26.1 & 26.0 &473.6& 781 &14208&10.3&50.3&308&1508 & 7.9 \\ 
\hline 
BP-6 & 13.4 & 13.4 & 473.6&401 & 14208&0.4&27.0&11&809 & 0.4 \\ 
\hline 
\end{tabular}
\caption{Same as Table~\ref{tab:OSSF-10TeV} for $\sqrt{s}=14\tev$ and an
  integrated luminosity of $30\fb^{-1}$. }
\label{tab:OSSF-14TeV}
\end{center}
\end{table}

Looking at the significance, we can now better see the importance of imposing
the Cut-3, i.e.\ the dilepton invariant mass should be in a window of $\pm
20\gev$ around $M_{A_H}$.  For instance, BP-4 has 2135 dilepton events for
$30\fb^{-1}$, but they are not coming from the decay $A_H \to l^+ l^-$ as
mentioned earlier, but instead from $Z$-decays. After Cut-2 we get a large
apparent statistical significance of $S_2/\sqrt{B_2} = 17.9$ (note that
$\sqrt{B_2} = 119.2$), but the huge reduction in the number of events after
Cut-3, from 2135 down to 56 events, tells us that these dileptons are not
coming from the decay of $A_H$, instead they come from an almost flat
background. Of course, this can be clearly seen by looking at the plot of the
dilepton invariant mass distribution in
Fig.~\ref{fig:inv_mass_dilepton_14TeV}. On the other hand, the Cut-3 removes
only about one third of the signal events for BP-3 and about half the events
for BP-1, BP-2 and BP-5. This indicates the presence of a peak around
$M_{A_H}$ for these benchmark points, see
Fig.~\ref{fig:inv_mass_dilepton_14TeV}. But note that for BP-4 even after
Cut-3, we still have $S_3/\sqrt{B_3} = 2.0$, but, of course, there will be no
peak in the dilepton distribution around $A_H$. Even for BP-6 we get after
Cut-2 a signal of $S_2/\sqrt{B_2} = 3.4$, but again they are not from the
dilepton decay of the $A_H$, as can be seen after Cut-3 is applied where we
get $S_3/\sqrt{B_3} = 0.4$.

We have seen in the previous Section~\ref{subsubsec:Dilepton_10-TeV} for the
LHC running at $\sqrt{s} = 10\tev$ and with $200\pb^{-1}$ integrated
luminosity that we had a clear dilepton signal over the SM
background for BP-1 and BP-3.

It is obvious from the last column in Table~\ref{tab:OSSF-14TeV} that with a
center of mass energy of $14\tev$ and an integrated luminosity of
$30\fb^{-1}$, we can now also cover BP-2 and BP-5. That means it will be
possible at these benchmark points to reconstruct the peak of $A_H$ in the
dilepton invariant mass distribution and to determine the mass $M_{A_H}$ and
the scale $f$. In particular with an integrated luminosity of $11.9\fb^{-1}$,
it will be possible to get $5\sigma$ statistical significance for BP-5 after
Cut-3. As a reminder, BP-5 has a rather heavy T-odd quark mass of $m_{q_H}
\sim 1\tev$. Furthermore we have chosen the difficult intermediate region with
$f=1\tev$, where the dilepton mode is not dominant. For lower $f$, the
discovery will be even easier for the same $m_{q_H}$ or, conversely, for lower
$f$, we get a larger reach in $m_{q_H}$, if we demand a $5\sigma$ signal with
$30\fb^{-1}$.

Again we should caution the reader about the numbers given in
Table~\ref{tab:OSSF-14TeV} for the SM background cross-sections and the number
of BG events after Cut-2 and in particular after Cut-3, because of the limited
statistics in the Monte-Carlo simulation. The total number of MC events in the OSSF channel we simulated (including all possible processes) is 
139429 after Cut-1. After the cut on $M_{eff}$ (Cut-2), we have 4170
events in our MC sample and after Cut-3 there remain 730 MC events in the
window of $\pm 20\gev$ around the $A_H$ mass (for the case $M_{A_H} =
66\gev$). Therefore, there is an intrinsic uncertainty of about 4\% on the
numbers for $B_3$ given in the Table~\ref{tab:OSSF-14TeV} and
correspondingly about 2\% uncertainty on the significance $S_3/\sqrt{B_3}$. In
particular, the required luminosity for a $5\sigma$ statistical significance
for BP-5 should be taken with a grain of salt. A much larger MC simulation to
pin down these numbers more precisely is again beyond the scope of the present
work. Note, however, that we have enough simulated events for the signal. For
instance, for BP-5, we have 12365 MC events after Cut-2 and 4880  MC
events after Cut-3.


\subsection{Four-lepton signal}

\subsubsection{LHC with $\sqrt{s} = 10\tev$}

The biggest SM background for the four-lepton signal, as defined in
Section~\ref{sec:event_selection}, arises from the production of $ZZ~+$~jets and the
subsequent fully leptonic decays.\footnote{For details on a method to normalize
the production rate of $ZZ$ from data see Ref.~\cite{Bruce}.}
For the LHC running at $\sqrt{s} = 10\tev$,
the corresponding cross-section after the basic cuts (Cut-1) is $13.78\fb$. It
is drastically reduced to $0.14\fb$ by the cut on the effective mass $M_{eff}
\geq 1\tev$ (Cut-2). The second largest SM background is from $t\bar{t}~+$~jets, but
after the basic cuts and the requirement that at least two OSSF leptons among the total four have an invariant mass in a window of $\pm
20\gev$ around $M_Z$, it is only $0.53\fb$ and it is completely removed by the effective mass cut.

In Table~\ref{tab:4l-10TeV} we list for all the benchmark points,
except BP-3, the cross-sections for the four-lepton signal for the LHC
running at a center of mass energy of $10\tev$, successively after the
basic cuts (Cut-1), Cut-2 on the effective mass and Cut-3, i.e.\ the condition
that the four-lepton invariant mass should be in a window of $\pm
20\gev$ around the mass of $A_H$, see Eq.~(\ref{M_ll_cut}). The total
SM background after Cut-2 and Cut-3 is also given. Note that for BP-3
with $f=500\gev$, the BR of $A_H$ into $Z^{}Z^{(*)}$ is essentially
zero, see Table~\ref{tab:A_H-Decay}, and therefore we have not
included that benchmark point in the table.

\begin{table}[htb]
\begin{center}
\begin{tabular}{|c|r|r|r|r|r|}
\hline
&\multicolumn{1}{c|}{Cut-1} &\multicolumn{1}{c|}{Cut-2} &
\multicolumn{1}{c|}{Cut-2} & \multicolumn{1}{c|}{Cut-3} &
\multicolumn{1}{c|}{Cut-3} \\ 
&\multicolumn{1}{c|}{S} & \multicolumn{1}{c|}{S} & \multicolumn{1}{c|}{BG}
&\multicolumn{1}{c|}{S} & \multicolumn{1}{c|}{BG}\\ 
&\multicolumn{1}{c|}{(fb)} & \multicolumn{1}{c|}{(fb)}  
&\multicolumn{1}{c|}{(fb)} & \multicolumn{1}{c|}{(fb)}
&\multicolumn{1}{c|}{(fb)} \\ 
\hline
BP-1 &21.70  &6.45&0.14&1.53& 0.003 \\
\hline
BP-2 & 1.21&1.07&0.14& 0.19& 0.003 \\
\hline
BP-4 &1.34 &1.26&0.14&0.33&0.010\\
\hline 
BP-5 & 0.23 & 0.23&0.14& 0.02& 0.003\\
\hline 
BP-6 & 0.18 & 0.18&0.14&0.04& 0.010\\
\hline 
\end{tabular}
\caption{Four-lepton signal cross-sections (S) for $\sqrt{s}=10\tev$ after the
  basic cuts (Cut-1), after Cut-2 and after Cut-3. For BP-3 there is no
  signal. The total SM background (BG) cross-section (mostly from $ZZ$) after
  Cut-2 and Cut-3 is also given. Note that we always
demand that among the four leptons, there is always at least one OSSF lepton
pair with its invariant mass being around $\pm 20 \gev$ of $M_Z$.}
\label{tab:4l-10TeV}
\end{center}
\end{table}

Unfortunately, although the four-lepton signal cross-sections are larger
than the SM background after Cut-2 for all the benchmark points considered in
the table, there will be always less than 10 signal events for an integrated
luminosity of $200\pb^{-1}$. Therefore we will not be able to see a clear
$A_H$ mass peak in the early stages of the LHC run.

\subsubsection{LHC with $\sqrt{s} = 14\tev$}

The biggest SM background for the four-lepton signal comes again from $ZZ~+$~ jets
production. For the LHC running at $\sqrt{s} = 14\tev$, the corresponding
cross-section after the basic cuts (Cut-1) is $18.79\fb$. It is drastically
reduced to $0.40\fb$ by the cut on the effective mass $M_{eff} \geq
1\tev$. The second largest SM background is from $t\bar{t}~+$~jets. After the basic
cuts and the requirement that at least two OSSF leptons among the total four have an invariant mass around $M_Z$, it is $0.26\fb$, but it is again completely removed by the effective mass cut.

\begin{figure}[t!]
\begin{center}
\centerline{\epsfig{file=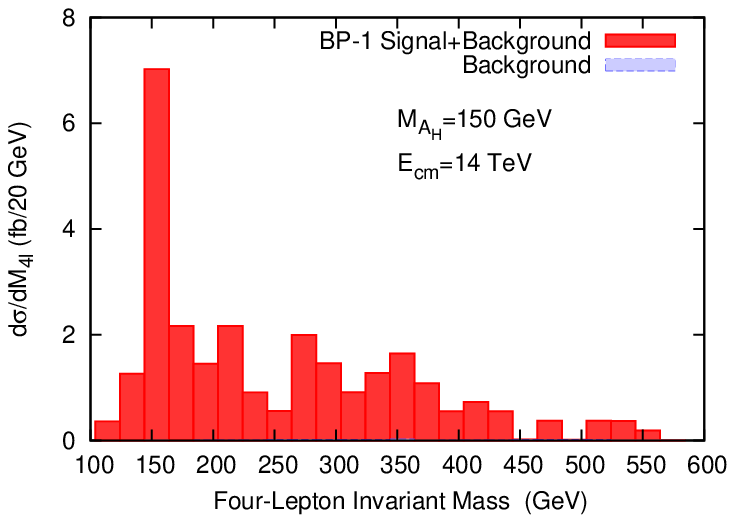,width=7.5cm,height=5.85cm,angle=-0}
\hskip 20pt \epsfig{file=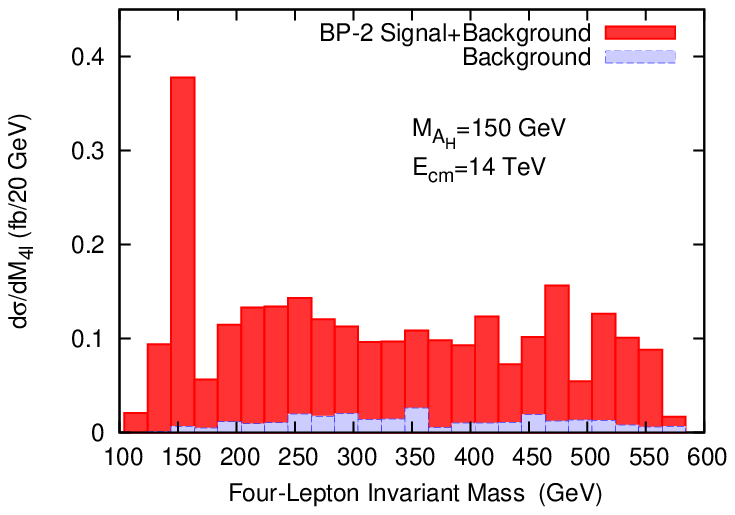,width=7.5cm,height=5.85cm,angle=-0}}
\centerline{\epsfig{file=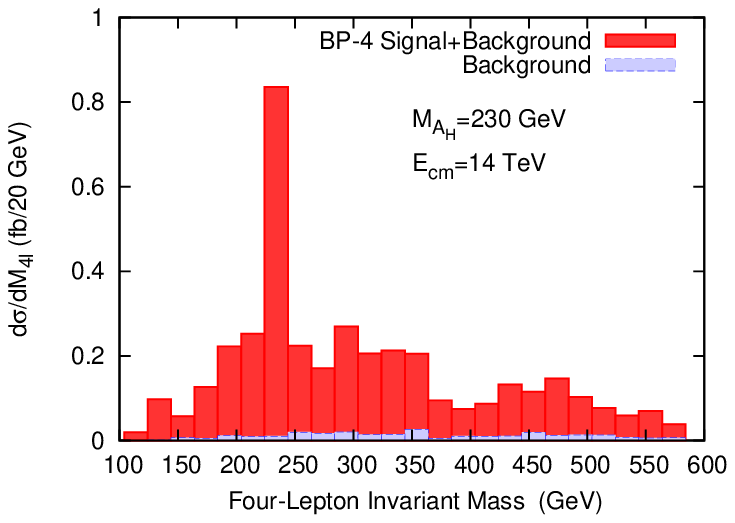,width=7.5cm,height=5.85cm,angle=-0}
\hskip 20pt \epsfig{file=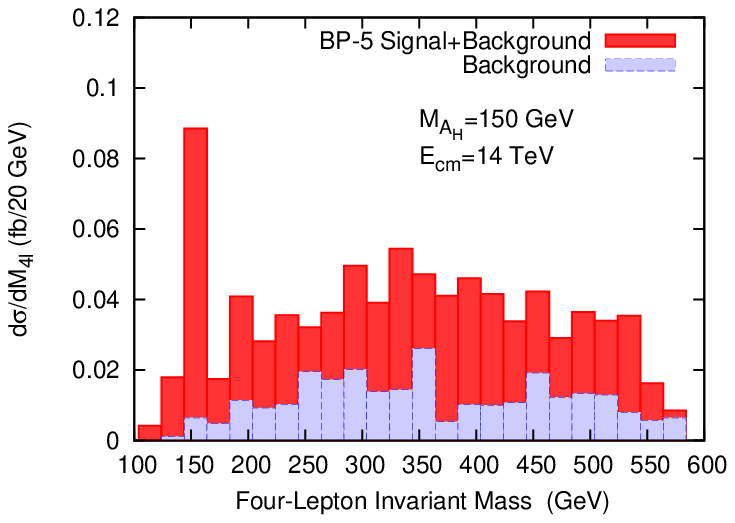,width=7.5cm,height=5.85cm,angle=-0}}
\centerline{\epsfig{file=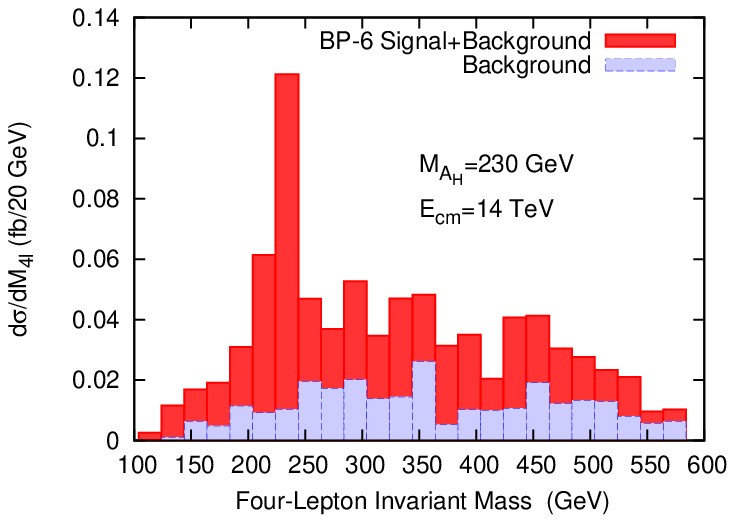,width=7.5cm,height=5.85cm,angle=-0}}
\caption{Four-lepton invariant mass distribution for all the benchmark points,
  except BP-3, where we do not expect any signal, after the basic cuts (Cut-1)
  and after the cut on $M_{eff} \geq 1\tev$ (Cut-2) for
  $\sqrt{s}=14\tev$.} 
\label{fig:4l-14TeV}
\end{center}
\end{figure}

In Fig.~\ref{fig:4l-14TeV} we show the four-lepton invariant mass
distributions for all the benchmark points, except BP-3, and the corresponding
SM background for the LHC running at $\sqrt{s} = 14\tev$ after the basics cuts
(Cut-1) and the cut of $M_{eff}$ (Cut-2).

As can be seen from Fig.~\ref{fig:4l-14TeV}, after the effective mass cut (Cut-2) the signal is larger than the SM background, except for BP-5 and BP-6.
But for all benchmark points a clear peak emerges at the mass of $A_H$. Note that we do not observe
any four-lepton events with an invariant mass of less than $100\gev$, which 
reflects the fact that we demand at least one OSSF lepton pair to
have an invariant mass around $M_Z$.

In Table~\ref{tab:4l-14TeV} we list for all the benchmark points, except BP-3,
the cross-sections for the four-lepton signal for the LHC running at a center
of mass energy of $14\tev$ after the basic cuts (Cut-1), after Cut-2 on the
effective mass and after Cut-3. The total SM background after Cut-2 and Cut-3
is also given. In addition, we give the number of events for an integrated
luminosity of $30\fb^{-1}$ after Cut-2 and after Cut-3.

\begin{table}[htb]
\begin{center}
\begin{tabular}{|c|r|r|c|r|c|r|r|r|r|}
\hline
&\multicolumn{1}{c|}{Cut-1} & \multicolumn{1}{c|}{Cut-2} &
\multicolumn{1}{c|}{Cut-2} & \multicolumn{1}{c|}{$S_2$} &
\multicolumn{1}{c|}{$B_2$} & \multicolumn{1}{c|}{Cut-3} &
\multicolumn{1}{c|}{Cut-3}&\multicolumn{1}{c|}{$S_3$} &
\multicolumn{1}{c|}{$B_3$}  \\ 
&\multicolumn{1}{c|}{[S]} & \multicolumn{1}{c|}{[S]} &
\multicolumn{1}{c|}{[BG]}& & & \multicolumn{1}{c|}{[S]} 
&\multicolumn{1}{c|}{[BG]}& & \\
&\multicolumn{1}{c|}{(fb)} &\multicolumn{1}{c|}{(fb)} &
\multicolumn{1}{c|}{(fb)} & & & \multicolumn{1}{c|}{(fb)} &
\multicolumn{1}{c|}{(fb)} &     &  \\   
\hline
BP-1 &58.85  & 28.80 & 0.40 &  864.0 & 12 & 9.18 &0.01
&275.4&0.3 \\ 
\hline
BP-2 &3.12   & 2.89  & 0.40 &  86.7  & 12 & 0.47 &0.01 &14.1
&0.3\\ 
\hline
BP-4 &4.25   & 4.02  & 0.40 &  120.6 & 12 & 1.12 &0.03 &33.6
&0.9 \\ 
\hline  
BP-5 & 0.79  & 0.79  & 0.40 &  23.7  & 12 & 0.10 &0.01 &3.0
&0.3  \\  
\hline 
BP-6 & 0.67  & 0.66  & 0.40 &  19.8  & 12 & 0.17 &0.03 &5.1
&0.9 \\  
\hline 
\end{tabular}
\caption{Same as Table~\ref{tab:4l-10TeV} for $\sqrt{s} = 14\tev$. The number
of signal and background events after Cut-2 with an integrated luminosity of
$30 \fb^{-1}$ are also given as $S_2$ and $B_2$, respectively. The
corresponding numbers after Cut-3 are denoted by $S_3$ and $B_3$.}
\label{tab:4l-14TeV}
\end{center}
\end{table}

We can see from the table that Cut-2, as for the dilepton signal, removes
about half of the four-lepton events for BP-1, but this is compensated by the
large parton-level cross section for $m_{q_H} \sim 400\gev$. The Cut-3 reduces
the signal by a factor of six for BP-2, by a factor of four for BP-4 and by a
factor of three for BP-1. 

It is obvious that with essentially no background, we have a very clear signal
and many events after Cut-3 for BP-1 and BP-4 with $30\fb^{-1}$ of integrated
luminosity. This should allow the reconstruction of the peak of the $A_H$
boson and the determination of the mass $M_{A_H}$. For BP-2 we have about 14
events after Cut-3, therefore the reconstruction of the peak might not be so
precise.

Recall that from the dilepton signature, we had a significant signal over the
background for BP-1 and BP-3 at $\sqrt{s} = 10\tev$ and with $200\pb^{-1}$ of
integrated luminosity. In addition, at $\sqrt{s} = 14\tev$ and with
$30\fb^{-1}$, we could also cover BP-2 and BP-5. Now, with four-lepton events,
we get a very clear signal after Cut-3 for BP-4 (34 events compared to about 1 background event) and, for an integrated luminosity of about $59\fb^{-1}$, we would get at least 10 signal events even for BP-6, with around 2 background events. Both of these benchmark points have $f=1500\gev$ and the branching fraction $A_H \to ZZ$ is $22.5\%$ and thus it is enhanced compared to BP-1 and BP-2, where for $f=1000\gev$ it is only about $11\%$, see Table~\ref{tab:A_H-Decay}. This allows us to use the four-lepton signal for discovery for BP-4 and, maybe, BP-6.

Of course, it would be a convincing cross-check on the LHT model with T-parity
violation, if one could see the $A_H$ peak in the dilepton and in the
four-lepton channel at the same mass. At least for BP-1 and BP-2 this will be
possible with the LHC running at $14\tev$ and with $30\fb^{-1}$. For BP-5, we
would need $100\fb^{-1}$ to get 10 four-lepton signal events after Cut-3.
Note that other New Physics models which have such a $Z^\prime$-type boson
like the $A_H$ might lead to a different pattern in the invariant mass
distributions or the relative number of events in the dilepton and four-lepton
channels might be very different from the ones in the LHT.

As suggested in Ref.~\cite{Zprime_to_ZZ}, looking at the angular distributions
of the two lepton pairs coming from $ZZ$-decays, one might be able to
determine whether the decay $A_H \to ZZ$ is really described by a vertex which
originates from the WZW-term. At least for BP-1 with 275 four-lepton events
after Cut-3, such an analysis seems to be feasible.


\section{Summary and Conclusions}
\label{sec:conclusions}

In this work we have analyzed dilepton and four-lepton events (here lepton
means electron or muon) at the LHC originating from the decay of the heavy
photon $A_H$ in the Littlest Higgs model with T-parity violation. These decays
of $A_H$, assumed to be the lightest T-odd particle, are induced by
T-violating couplings from the Wess-Zumino-Witten anomaly term.

Since the WZW term reproduces, within the EFT, the chiral anomalies in the UV
completion of the LHT, its prefactor $N/48\pi^2$ corresponds to a one-loop
effect. Furthermore, because of gauge invariance, the actual coupling of $A_H$
to SM gauge bosons has an additional suppression factor of $v^2/f^2$. For
larger masses $M_{A_H} > 150 \gev$, $A_H$ predominantly decays into $W
W^{(*)}$ and $Z Z^{(*)}$. On the other hand, for smaller masses, loop-induced
decays into SM fermions are possible.  The corresponding one-loop diagrams in
the EFT are, however, UV divergent and one needs counterterms with {\it a
priori} unknown coefficients. Following Ref.~\cite{Freitas_Schwaller_Wyler}
these coefficients have been fixed by naive dimensional analysis. The
couplings of $A_H$ to SM fermions are then effectively of the size of two-loop
effects.

Due to the tiny coupling of $A_H$ to SM particles, its direct production at
$e^+ e^-$ or hadron colliders only has a cross-section of the order of
$10^{-6}\pb$. On the other hand, the production of the other T-odd particles
is not affected by the presence of the WZW term. Therefore, these particles
are still pair-produced and cascade decay down to $A_H$ by the T-conserving
interactions in the LHT and finally the two $A_H$'s decay promptly in the
detector.  The crucial observation, already made in Ref.~\cite{Zprime_to_ZZ},
is that summing all production processes of heavy T-odd quark pairs leads to a
sizeable cross-section at the LHC, of the order of several pb, and this
corresponds to a lower bound on the production cross-section of $A_H$ pairs.

We have studied the dilepton and four-lepton signals for six benchmark points,
see Table~\ref{tab:BPs}, which have different values for the heavy quark mass
$m_{q_H} \sim 400, 700, 1000\gev$ and different values for the mass of the
heavy photon $M_{A_H} = 66, 150, 230 \gev$ ($f= 500, 1000, 1500\gev$). The
values of the heavy quark mass essentially determine the parton-level
pair-production cross-section via strong interaction processes, although the
effects of electroweak contributions from $t$-channel exchanges of $A_H$ and
$Z_H$ can be very important and even dominate for low values of $f$, i.e.\
light $A_H$ and $Z_H$. On the other hand, the mass of $A_H$ determines the
expected signal because of the different branching ratios into leptons or
$ZZ$, see Table~\ref{tab:A_H-Decay} . For low and intermediate masses of
$A_H$, the dilepton decays are sizeable, whereas for $M_{A_H} = 230\gev$ the
decay into $ZZ$ and then into four-leptons is relevant.

We have studied the case of the LHC running at 
a center of mass energy of $\sqrt{s} = 10\tev$ with a modest integrated
luminosity of $200\pb^{-1}$ and the case of $\sqrt{s} = 14\tev$ with an
integrated luminosity of $30\fb^{-1}$.

In order to reduce the SM background from $Z/\gamma^{*}$ and $t\bar{t}~+$~jets for the dilepton signal and from $ZZ~+$~jets for the four-lepton signal, we have imposed a large cut on the effective mass of the events of $M_{eff} > 1\tev$, since in general $M_{eff}$ approximately peaks at the sum of the masses of the
initially produced particles. Essentially, this cut removes a considerable fraction of the SM backgrounds, except for processes with multiple additional hard jets, which have a long tail in the effective mass
distribution (see Figure~\ref{fig:Meff}). On the other hand, there are many
sources of leptons in the decay cascades leading to $A_H$ and in general we
also get many events with leptons which do not originate from the decay of
$A_H$. We have reduced the corresponding background from within the LHT by
imposing the condition that the invariant mass of dileptons or four leptons
should lie in a window of $\pm 20\gev$ around $M_{A_H}$.

For the {\it dilepton signal}, the main conclusion is that for regions of the
parameter space where either the T-odd quarks are relatively light, $m_{q_H}
\sim 400\gev$ (BP-1 with $f= 1000\gev$) or the scale $f$ is rather low,
$f=500\gev$ (BP-3 with $m_{q_H} \sim 700\gev$), we get after all the cuts a
clear signal above the background for the early run of the LHC with center of
mass energy of $10\tev$ and integrated luminosity of $200\pb^{-1}$. More
details can be found in Table~\ref{tab:OSSF-10TeV}. For the LHC with $\sqrt{s}
= 14\tev$ and $30\fb^{-1}$ luminosity, also BP-2 ($m_{q_H} \sim 700\gev$,
$M_{A_H} = 150\gev$) and BP-5 ($m_{q_H} \sim 1000\gev$, $M_{A_H} = 150\gev$)
yield a significant signal with $S/\sqrt{B} = 42.9$ for the former and
$S/\sqrt{B} = 7.9$ for the latter benchmark point, see
Table~\ref{tab:OSSF-14TeV} for details.

The {\it four-lepton channel} is very clean and the signal cross-sections are
larger than the SM backgrounds after all the cuts, with the exception of
BP-3 with small $f=500\gev$, where we do not expect a four-lepton
signal. Unfortunately, for the LHC running at $10\tev$ and with $200\pb^{-1}$
of integrated luminosity, we always get less than 10 signal events. For
$\sqrt{s} = 14\tev$ and $30\fb^{-1}$, the background is again negligible ($<
0.9$~events) and we can easily cover again BP-1 and BP-2. In addition, we now
also get a clear signal for BP-4 ($m_{q_H} \sim 700\gev$, $M_{A_H} =
230\gev$). We would need $59\fb^{-1}$ to get 10 signal events for BP-6
($m_{q_H} \sim 1000\gev$, $M_{A_H} = 230\gev$). Note that these BP's with
large $f=1500\gev$ can only be covered in the four-lepton channel. Details can
be found in Table~\ref{tab:4l-14TeV}.

Therefore, with the LHC running at $14\tev$ and an integrated luminosity of
$30\fb^{-1}$, we can cover with the dilepton and / or the four-lepton signal a
large part of the typical parameter space of the LHT with values of $f$ up to
$1500\gev$ and with T-odd quark masses up to about $1000\gev$.  In general, a
clear peak emerges at $M_{A_H}$, if one plots the invariant mass distributions
for dileptons, see Fig.~\ref{fig:inv_mass_dilepton_10TeV} for the LHC running
at $10\tev$, and Fig.~\ref{fig:inv_mass_dilepton_14TeV} for $\sqrt{s} =
14\tev$. The four-lepton invariant mass distribution for the LHC at $14\tev$
is shown in Fig.~\ref{fig:4l-14TeV}. For all the studied benchmark points we
have enough signal events after all the cuts, therefore it should be easy to
reconstruct the mass peak of $A_H$, maybe with the exception of BP-6.

Note that the reconstruction of the peak and the measurement of $M_{A_H}$
directly determines the symmetry breaking scale $f$ in the LHT, which is one
of the fundamental parameters of any Little Higgs model. Together with the
$M_{eff}$ distribution which peaks around $2 m_{q_H}$, this would then allow
a rough determination of the parameter $\kappa_q$ as well.

Of course, it would also be an important cross-check on the LHT model with
T-parity violation, if we could see the $A_H$ peak in both the dilepton and
the four-lepton channel at the same mass. This, including the ratio of events
in the two channels, distinguishes the case of $A_H$ in the LHT from other
models of New Physics, like most $Z^\prime$ models. At least for benchmark
points with intermediate values of $f = 1000\gev$, which yield both enough
dilepton and four-lepton events, this could be achieved. For BP-1 and BP-2
this will be possible for the LHC running at $14\tev$ and with
$30\fb^{-1}$. For BP-5 we would need $100\fb^{-1}$ to get 10 four-lepton
events (with essentially very low background).

Finally, we should stress again that the WZW term offers a unique window into
the UV completion of the LHT. The parameter $N$ determines the total decay
width of $A_H$, but since the width is only a few eV, it cannot be measured
experimentally. Furthermore, in the decay branching ratios, the factor $N$
drops out. On the other hand, the finite parts of the counterterms needed to
renormalize the one-loop diagrams in the EFT are determined by the underlying
theory. In principle, measuring the branching ratios of $A_H$ would therefore
yield information on the more fundamental theory. As briefly discussed, if
these branching ratios of the decays into leptons or quarks differ
substantially from those of the $Z$-boson, one might even be able to see a
signal from $A_H$ with a mass close to $M_Z$.

\vspace{0.3cm}
\noindent
{\bf Note added.} After this manuscript was first submitted, the   LHC schedule was revised, targetting a run at 7 TeV with about $1-2 ~\fb^{-1}$ luminosity, followed by a direct upgrade to 14 TeV. Since the 14 TeV run maximizes our reach in the parameter space, we have presented the results corresponding to this energy in detail. However, we have also retained the results for 10 TeV.  

\section*{Acknowledgments} 

S.M.\ would like to thank Sanjoy Biswas and Nishita Desai for many useful
discussions and technical help. We also thank V.\ Ravindran for valuable discussions and Bruce Mellado for an important
comment on the manuscript. This work was partially supported by funding
available from the Department of Atomic Energy, Government of India, for the
Regional Centre for Accelerator-based Particle Physics (RECAPP),
Harish-Chandra Research Institute.  Computational work for this study was
partially carried out at the cluster computing facility in the Harish-Chandra
Research Institute (http:/$\!$/cluster.hri.res.in).


\end{document}